\documentclass{jkas}

\def\year{2019} 
\def\month{October} 
\def\volume{52} 
\def\issue{5} 
\def\beginpage{145} 
\def\endpage{158} 
\def\received{May 14, 2019} 
\def\accepted{July 16, 2019} 
\date{Received \received; accepted \accepted}

\def\etal{{et al.}}
\def\lesssim{\mathrel{\hbox{\rlap{\hbox{\lower4pt\hbox{$\sim$}}}\hbox{$<$}}}}
\def\gtrsim{\mathrel{\hbox{\rlap{\hbox{\lower4pt\hbox{$\sim$}}}\hbox{$>$}}}}
\def\apj{ApJ}

\def\aj{AJ}
\def\aap{A\&\hskip-1pt A}

\def\mn{MNRAS}
\def\pasp{PASP}
\def\araa{ARA\&\hskip-1pt A}

\def\kas{JKAS}

\usepackage{flushend} 
\usepackage{microtype}
\usepackage[rightcaption]{sidecap}

\usepackage[colorlinks,
            pdftex, 
            breaklinks,
            linkcolor=blue,
            urlcolor=blue,
            anchorcolor=blue,
            menucolor=blue,
            citecolor=blue]{hyperref}

\title{
207 New Open Star Clusters within 1 kpc from Gaia Data Release 2
}

\author[1]{Gyuheon~Sim}
\author[2,3]{Sang~Hyun~Lee}
\author[4]{Hong~Bae~Ann}
\author[1]{Seunghyeon~Kim}

\affil[1]{Ulsan Science High School, Ulsan, 44902, Korea; \email{2017000049@ushs.hs.kr, 2017000016@ushs.hs.kr}}
\affil[2]{Korea Astronomy and Space Science Institute, Daejeon 34055, Korea; \email{shlee@ksai.re.kr}}
\affil[3]{Department of Physics, University of Ulsan, Ulsan 44610, Korea}
\affil[4]{Pusan National University, Busan 46241, Korea; \email{hbann@pusan.ac.kr}}

\def\corrauthor{
H. B. Ann
}

\def\runningauthor{
Sim et al.
}

\def\runningtitle{
207 New Open Star Clusters within 1 kpc from Gaia DR2
}

\def\keywords{
open clusters and associations: general --- catalogs --- methods: data analysis
}

\def\abstracttext{
We conducted a survey of open clusters within 1 kpc from the Sun using the
astrometric and photometric data of the \emph{Gaia} Data Release 2. We found 655 cluster candidates
by visual inspection of the stellar distributions in proper motion space and spatial distributions in $l-b$ space.
All of the 655 cluster candidates have a well defined main-sequence except for two candidates if we consider that the main sequence of very young clusters is somewhat broad due to differential extinction. Cross-matching of our 653 open clusters with known open clusters in various catalogs resulted
in 207 new open clusters. We present the physical properties of the newly discovered open
clusters.
The majority of the newly discovered open clusters are of young to intermediate age and have less than $\sim$50 member stars.
}

\begin{document}
\jkashead 


\section{Introduction\label{sec:intro}}

Open clusters are local concentrations of stars with a common sense of motion \citep{tru31}. They are thought to have been formed in the same
nebulae.  \citet{tru30} classified open clusters into four
types according to central concentration,
luminosity range of member stars, and number of cluster members. Open clusters are
good tracers of the Galactic structure \citep{jan82, jan94} as well as test beds for stellar
evolution theory \citep{san57, mer81} and dynamics of stellar systems \citep{ter87}.
However, owing to severe contamination of cluster images with field stars, it is challenging to determine
their sizes and morphologies. Evaporation of low mass stars as part of the dynamical evolution of star clusters
makes it difficult to distinguish member stars in the outer regions of clusters from field stars even if kinematic data are available \citep{fri08}.

Multiple catalogs of open clusters are available. The Lund Catalogue
of Open Cluster Data \citep{lyn87} lists about 1200 open clusters which have been intensively
used as targets of photometric surveys such as that of \citet{ann99}.
\citet{dia02} compiled 1537 open clusters, updating
previous catalogs including those of \citet{lyn87} and \citet{mer95}.
After publications of the DALM catalog by \citet{dia02}, multiple studies aimed at discovering
new open clusters by analyzing stellar catalogs such as the All-Sky Compiled Catalogue (ASCC-2.5) \citep{kha01, ros10}. \citet{kha05} discovered 109 new open clusters using the ASCC-2.5 while
\citet{sch15} discovered 63 new open clusters using the PPXML catalog \citep{ros10}. Both catalogs
are all sky catalogs that provide positions and proper motions in the International Celestial Reference
System (ICRS) together with the near-infrared magnitudes from 2MASS \citep{skr06} for
399 million objects. These new discoveries were a by-product of constructing the Catalogue of Open
Cluster Data (COCD) from ASCC-2.5 and the global Milky Way Star Clusters (MWSC) survey which
compiled $\sim$3000 clusters from PPXML. The work of \citet{mer95} has evolved into
the WEBDA database.\footnote{\url{https://webda.physics.muni.cz}}
DALM is continuously updated and lists about 2000 open clusters to date.

Until recently, visual inspections of photographic plates or CCD images were the most widely used method
to search for open clusters. However, there are two notable movements in this field: extension to
other wavelengths, and computer-aided search. Infrared observations have been quite successful in finding embedded clusters \citep{cam15a, cam15b, ryu18, oli18}. Embedded clusters are groups of very young stars that are partially or fully obscured by
interstellar gas and dust. Computer-aided search applies automated search algorithms to large stellar databases. \citet{kop17} detected a new star cluster by analyzing overdensity in spatial distributions of all sources in the \emph{Gaia} Data Release 1 \citep{gai16b}. More elaborated algorithms such as UPMASK \citep{kro14} were applied to the \emph{Gaia} data with notable success \citep{cas18, can19}.

\begin{table}[t!]
  \caption{Size of search area.}
  \label{tab1}
\setlength{\tabcolsep}{4pt}
\centering
	\begin{tabular}{cccccc}
		\toprule
		$|b|( ^{\circ})$ & $<$15 & 15--21 & 21--30 & 30--80 & 80--90 \\
		\midrule
		Area & $2^{\circ}\times2^{\circ}$ & $3^{\circ}\times3^{\circ}$ & $9^{\circ}\times9^{\circ}$ & $10^{\circ}\times10^{\circ}$ & $90^{\circ}\times10^{\circ}$ \\
		\bottomrule
	\end{tabular}
\end{table}

The \emph{Gaia} mission \citep{gai16a} has made available highly precise astrometric measurements of more than one billion stars along with precise magnitudes in three passbands -- $G$ (330~nm--1050~nm), $G_{\rm{BP}}$ (330~nm--680~nm), and $G_{\rm{RP}}$ (630~nm--1050~nm).
The second release of \emph{Gaia} data (\emph{Gaia} DR2) provides celestial coordinates, proper motions and parallaxes
of about 1.3 billion stars along with photometric magnitudes \citep{gai18}.
\emph{Gaia} DR2 data have been used to find new open clusters in multi-dimensional
parameter spaces. \citet{cas18} and \citet{can19} found 31 and 41 open clusters from \emph{Gaia} DR2 using the cluster-finding algorithms
DBSCAN and UPMASK, respectively. \citet{can19} utilized a Gaussian mixture
model (GMM) along with UPMASK to detect star clusters. Application of
UPMASK to the astrometric data of \emph{Gaia} DR2 for 3328 open clusters compiled from various catalogs
in the literature, including those of \citet{fro06}, \citet{kha13}, and \citet{dia02}, yielded the identification of 1229
open clusters \citep{can18}.

Due to dynamical evolution, open clusters can change in morphology and size. They
also can be disrupted by encounters with giant molecular clouds \citep{gie06} and
flattened by Galactic tidal forces \citep{ter87}. Elongated morphologies have been reported for some clusters such as the Hyades \citep{oor79} and Pleiades \citep{rab98}. The Galactic tidal field can stretch clusters towards the Galactic center. In addition, elongation is frequent in
young open clusters \citep{che04}, indicating a link to cluster formation processes. Numerical simulations explicitly showed the role of dynamical
evolution for elongation \citep{kha09b}.  Tidal tails have been predicted to be aligned with the Galactic orbit \citep{chu10}. \citet{lee13}
found the weak signature of a tidal tail in the open cluster NGC~1245 from a statistical
analysis of deep CCD photometry. Recently, observational evidence of the Hyades tidal
tails was reported \citep{ros19}.

The aim of this study is to find new open clusters within 1 kpc from the Sun by visual inspection using multi-dimensional data of \emph{Gaia} DR2. Recent discoveries of open clusters from \emph{Gaia} DR2 were made through automated algorithms. However, most open clusters have been found visually.  We expect that visual inspection of clustering patterns in proper motion space and spatial distribution of stars will discover many open clusters not yet identified. In addition, the precision of the \emph{Gaia} astrometric data greatly reduces the difficulties in distinguishing cluster stars from background field stars. Since we restricted our survey to the stars within 1 kpc, we reach down to the faint main-sequence stars which are required to examine the outer regions of open clusters. We anticipate that the information of the outer structures of open clusters will greatly help us to broaden our view of the dynamical evolution of open clusters, field star formation, and structure of the Milky Way.

This paper is organized as follows. In Section~\ref{sec:data}, we explain the \emph{Gaia} DR2 data and the selection criteria used to extract the \emph{Gaia} DR2 data while minimizing field star contamination. In Section~\ref{sec:method}, we describe how we found clusters visually, validated clusters with the PARSEC set of metallicity $Z=0.02$ isochrones \citep{bre12} and derived their parameters quantitatively. In Section~\ref{sec:results}, we present our catalog of newly found clusters with a brief description of the properties of the new clusters. Finally, in Section~\ref{sec:sumdis}, we summarize and discuss our results.

\begin{figure}[!t]
	\includegraphics[width=1\columnwidth]{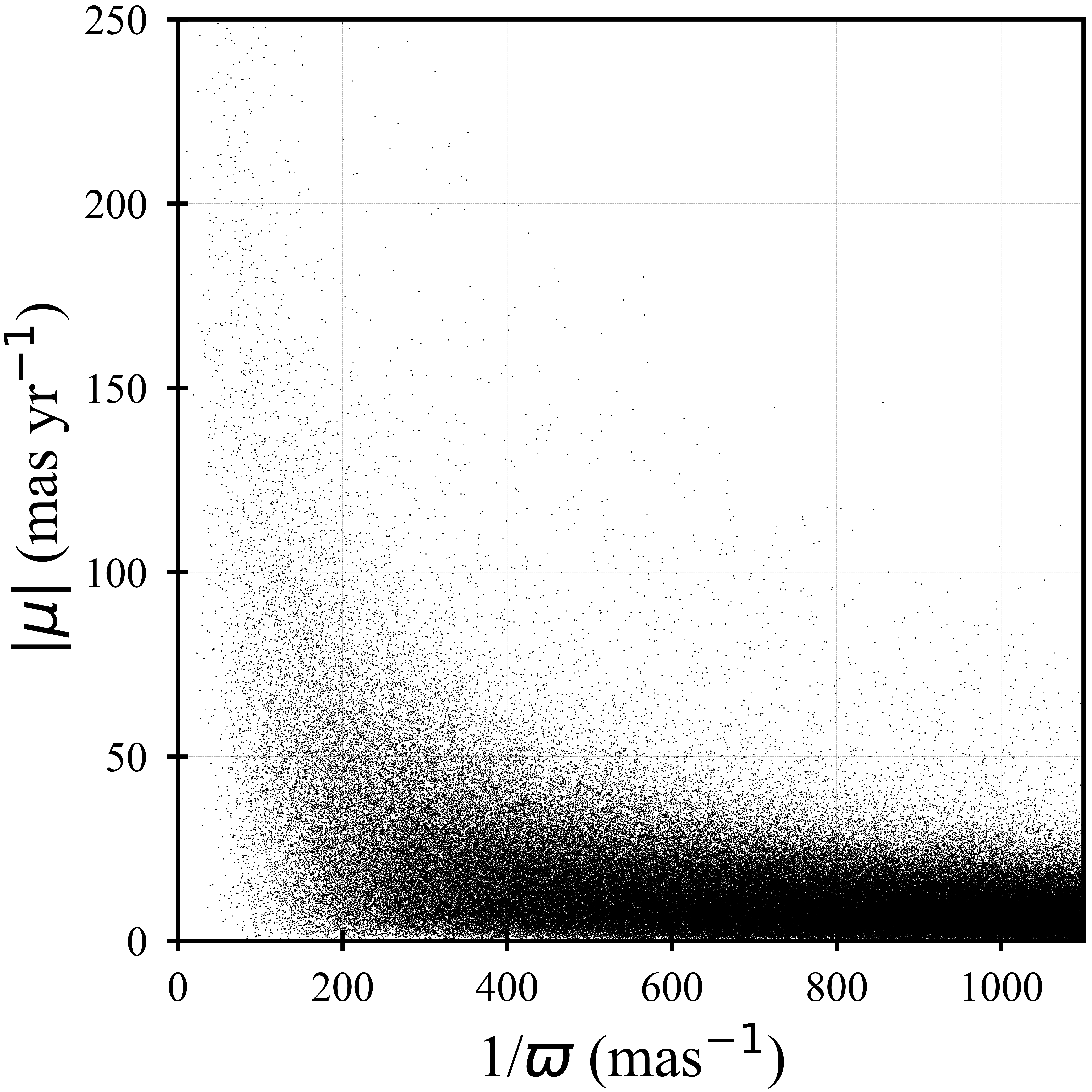}
	\caption{Proper motion vs. parallax of sample stars.}
      \label{ref:pm_pi}
\end{figure}

\section{Data}
\label{sec:data}

The \emph{Gaia} DR2 is based on data collected during the first 22 months of the nominal mission
lifetime. The biggest difference between DR1 and DR2 is that DR2 provides photometric data suitable for constructing color-magnitude diagrams (CMD) of clusters. We used the five astrometric parameters Galactic longitude $l$, Galactic latitude $b$, proper motion in right ascension $\mu_{\alpha}cos\delta$ (with $\delta$ being the declination), proper motion in declination $\mu_{\delta}$, parallax angle $\varpi$, and the three \emph{Gaia} photometric magnitudes, $G$, $G_{\rm{BP}}$, and $G_{\rm{RP}}$ to search for open clusters.
In general, the typical uncertainty of the astrometric measurements increases as the target brightness decreases. The typical uncertainties in position and parallax are
$\sim$0.04~mas for bright sources ($G<14$~mag) and 0.7~mas at $G=20$~mag \citep{lin18}. Along with these typical uncertainties, there is a zero-point offset ($-0.029$~mas) in the \emph{Gaia} parallaxes \citep{lin18}. We corrected our cluster distances for the zero-point offset.  Another source of uncertainty is an offset in the $G_{\rm{BP}}$ and $G_{\rm{RP}}$ magnitudes at the faint end ($G>19$) of the \emph{Gaia} data \citep{gai18}.

As described in \citet{gai18}, there are some sources that have spurious parallaxes.
Most of them are faint sources and concentrated in dense regions along the Galactic plane and
towards the bulge of the Galaxy \citep{lin18}. Since the uncertainty in a parallax propagates into the
estimates for distance-related parameters, it is better to exclude faint stars from the sample and to keep in mind the possibility of spurious parallaxes for stars in dense regions.
By considering the uncertainties in the five astrometric parameters of \emph{Gaia} DR2,
we selected stars brighter than $G=18$~mag with parallaxes $\varpi > 0.833$~mas.
For our magnitude cut, the errors in position and parallax are about 0.15~mas, errors in proper motion are about 0.3~mas\,yr$^{-1}$. Errors of this magnitude have little impact on the identification of
clusters in the multi-dimensional parameter space described below as they correspond to less than $\sim10\%$ of the mean proper motions and parallaxes of candidate clusters, respectively. A brightness cut $G < 18$ is necessary because large errors in positions
and parallaxes of faint stars make the visual detection of sparse clusters difficult. Figure~\ref{ref:pm_pi} presents the relationship between proper motion
and parallax, showing that most of the sample stars have proper motions less than
25~mas\,yr$^{-1}$. As expected, proper motions are smaller for more distant stars; there are some high proper motion stars most of which are located within $\sim$200~pc
from the Sun. These high proper motion stars are supposed to be high velocity stars.
Therefore, we expect that most clusters have proper motions less than 25~mas\,yr$^{-1}$. For nearby clusters, we should considered a larger range of proper motions.

\section{Method}
\label{sec:method}

Our process of searching open clusters from the \emph{Gaia} DR2 is divided into
two steps. The first step is the visual search for cluster candidates by looking at the multi-dimensional information of
\emph{Gaia} DR2 (proper motion, position, and parallax).
The second step is the verification of candidate clusters by comparing their CMDs with theoretical isochrones.
During the search, physical parameters of clusters
such as cluster radius were derived.

\subsection{Visual Inspection}

We divided the whole sky into 3628 areas for visual inspection of cluster candidates.
The size of the search area varied with  Galactic latitude ($b$). Each area was selected to have about $10^5$
stars; we set this upper limit for the number of stars in each area to reduce field
star contamination and to increase the contrast between field stars and cluster members.
Details are given in Table~\ref{tab1}.

\begin{figure}[!t]
	\includegraphics[width=1\columnwidth]{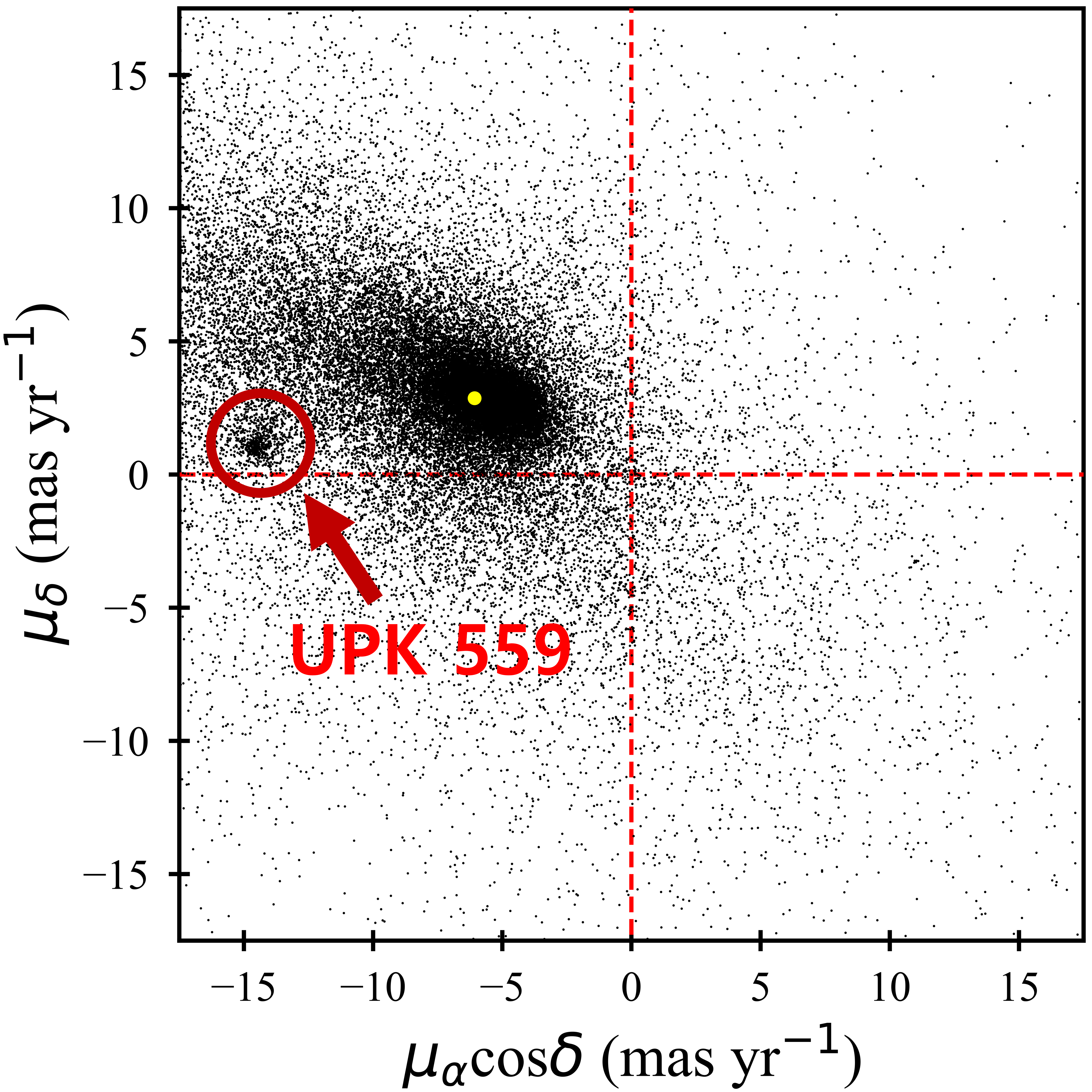}
	\caption{Proper motions of a sample of 47\,578 stars located within $2^{\circ}$ from $l=285^{\circ}$ and $b=0^{\circ}$.}
\label{fig:pm}
\end{figure}

We first checked the proper motions of stars in each area. The proper motion
distribution varies locally and cluster stars are ``buried'' in field stars (with few exceptions such as the Hyades and Pleiades). We therefore first identify the center of the
field star distribution in proper motion space, and then search for off-center local concentrations corresponding to stars of a prospective cluster. Figure~\ref{fig:pm} shows an example for a set of field stars centered at $\mu_{\alpha}\cos\delta=-6.08$~mas\,yr$^{-1}$ and $\mu_{\delta}=2.87$~mas\,yr$^{-1}$.

In a given search area, we first calculated the average
proper motion of all stars within 30~mas\,yr$^{-1}$ from $(0,0)$. We then re-calculated the average proper motion using all stars located within 30~mas\,yr$^{-1}$ from the newly found kinematic center; we repeated this procedure until it converged. Our approach reduces the impact of high-proper motion stars. After identification of the kinematic center of the field stars, we visually searched for off-center local concentrations of stars within a radius of 25~mas\,yr$^{-1}$. The search radius was obtained by trial and error, guided by the proper motion distribution as
a function of parallax shown in Figure~\ref{ref:pm_pi}.

\begin{figure}[!t]
	\includegraphics[width=1\columnwidth]{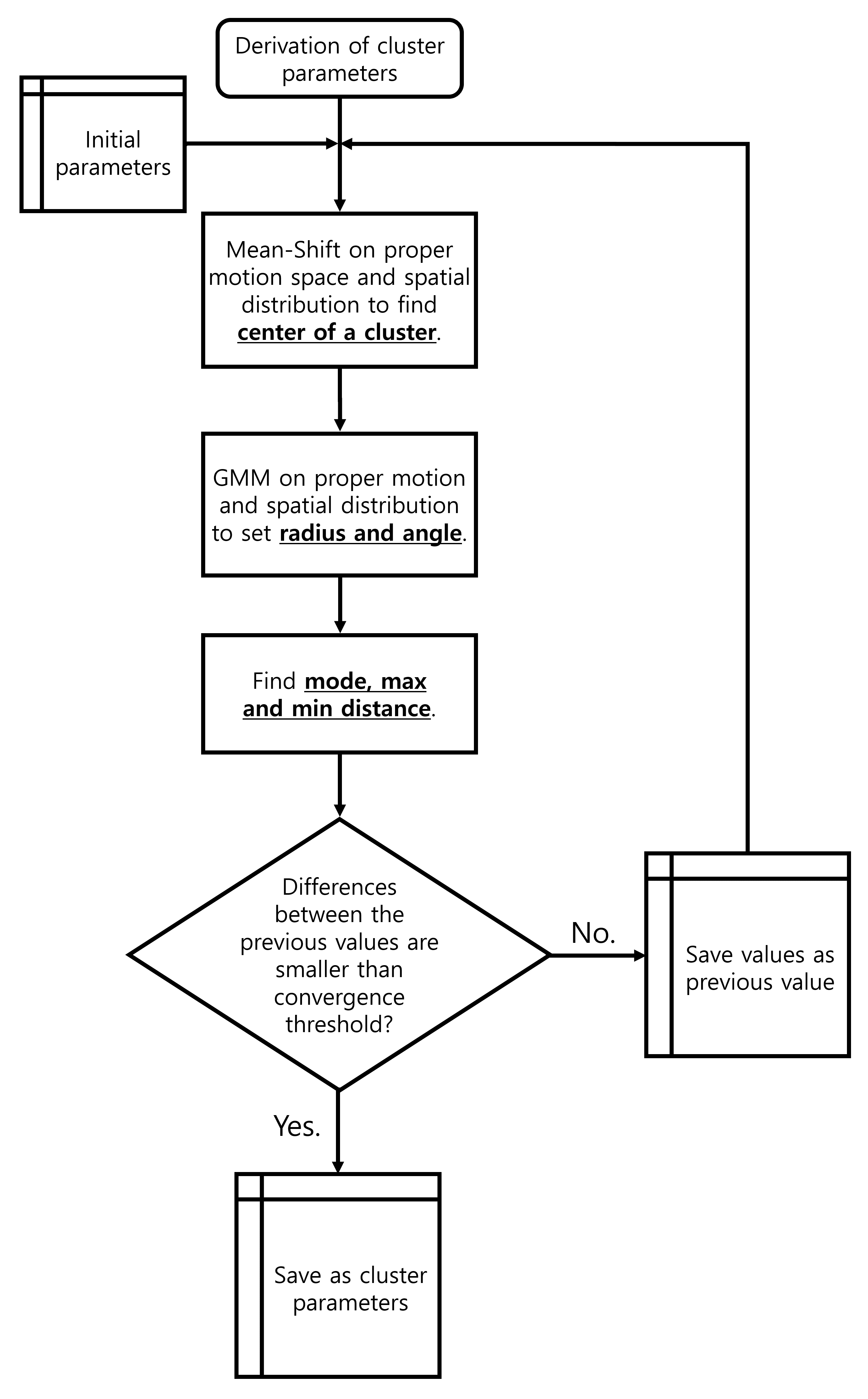}
	\caption{Iterative derivation of the physical parameters of cluster candidates identified from visual inspection. }
      \label{ref:flowchart}
\end{figure}

After identifying a candidate cluster in proper motion space, we obtained the (absolute) proper motions of its member stars and determined their spatial distribution in $l-b$ space.
In case of clusters with very large proper motions, such as the Hyades cluster, we included stars with proper motions in the range $\pm150$~mas\,yr$^{-1}$. We then checked the distribution of their parallaxes for a significant peak caused by a concentration of stars at a certain distance, as required for a physical cluster.  If the parallax distribution indeed showed such a
peak, we determined the most probable range of parallaxes wherein
the clusters stars are likely to be located. After visual inspection of the whole area, we inspected again all the selected cluster candidates to check for ambiguities and duplications. The final number of cluster candidates is 655. The cluster parameters determined by visual inspection were
used as initial values for the further analysis described below.

\subsection{Derivation of Cluster Parameters}

Since the cluster parameters obtained from visual inspection can only be rough estimates,
we further analyzed our cluster candidates using GMM and mean-shift algorithms.
For each cluster, we obtained its center and radius in proper motion and $l-b$ space, respectively, its distance from the mode
of the distribution of the parallaxes of the cluster stars, and upper and lower limits on the cluster distance, $d_{max}$ and $d_{min}$.

The covariance and Gaussian center provided by a GMM were used to derive the size
and shape of a cluster in terms of axis lengths, position angle (orientation on the sky), and ellipticity. For the GMM analysis, we used the Python library \texttt{Scikit-Learn} \citep{ped11}. We set the convergence
threshold to 0.001 and the mixture component to 1. The mean-shift algorithm was used to find the center of a density distribution. For GMM analysis in $l-b$ space, we weighted the data with the inverse distance from the
center. The weights were given by $w_{i}=(|r_{i} - 2r|/2r)^{p}$ where $r_{i}$ is the distance of a star $i$ from the center,
${r}$ is the cluster radius and $p$ is a power index that depends on ${r}$. Here, $r$
is defined as the radius in $l-b$ space ($r_{\rm{GMM}}$).

We summarize our analysis procedure in Figure~\ref{ref:flowchart}.
We derived the center and radius of a cluster in proper motion and $l-b$ space iteratively
using the parameters obtained from the visual inspection as initial values. When we
derived the centers of proper motion distributions by mean-shift,
we included all stars that are located within the distance range allowed for the cluster. Similarly, in $l-b$ space, we included all stars
that satisfy reasonable boundary conditions on proper motion distribution and distance.
Some clusters, such as small clusters in the Orion nebula, are located close to each other and
have almost the same proper motions. Clusters with similar proper motions happened to be added together by our analysis routines, allowing for an abnormal growth of search
radius which hindered a proper determination of the cluster center in proper motion space. We therefore limit the search
radius to three quarters of the radius of the stellar concentration in proper motion space determined by visual inspection.
We also limit the search radius in $l-b$ space to the radius determined by visual inspection.

Once cluster parameters in proper motion space were determined from mean-shift analysis,
we derived updated values for the cluster parameters in $l-b$ space using the stars that
satisfy the criteria for cluster distance and proper motion parameters. We applied ellipse fitting using GMM to determine the shape of the spatial distribution of cluster stars, using the stars
that are still compatible with being cluster members after updating the values for distances and range of proper motions. We derived the semi-major axis, semi-minor axis, and the angle between the major axis and the Galactic plane. We applied GMM to the proper motion data, assuming that cluster stars are located within  1$\sigma$ of the Gaussian model.
We defined a radius $r_{\rm{GMM}}$ as the radius for which the enclosed area is the same as the area of the ellipse derived by GMM. We determined the mode of the distribution of parallaxes by fitting a Gaussian kernel to the parallax distribution of stars that have proper motions allowed for the cluster and are located within the ellipse defined in the spatial distribution. The cluster distance range is given by the full width at half maximum (FWHM) of the Gaussian kernel.

\begin{figure}[!t]
  \centering
  \includegraphics[width=1\columnwidth]{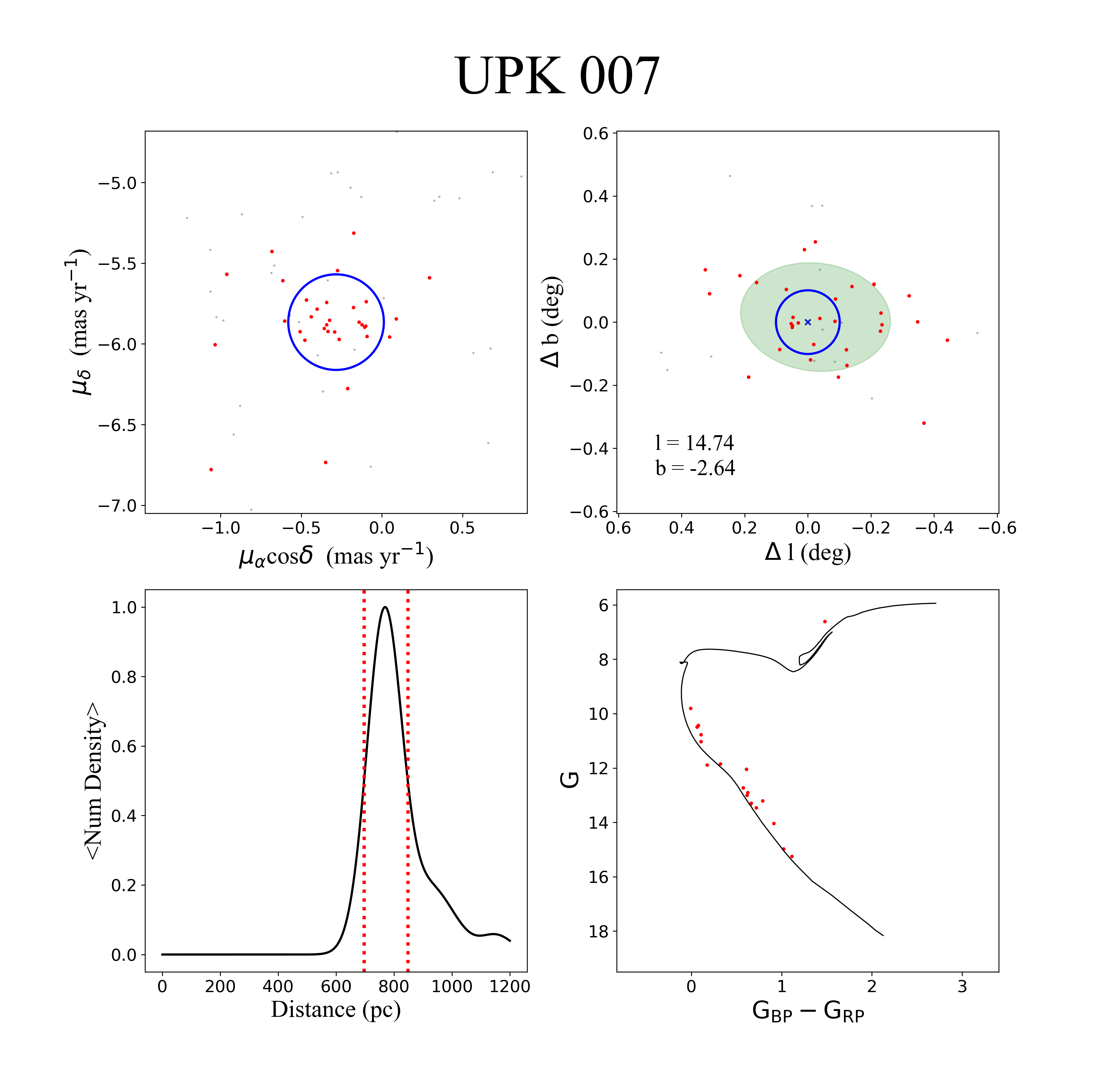}
  \caption{Properties of cluster candidate UPK~007. \emph{Top left:} Proper motion space. \emph{Top right:} $l-b$ space. Blue circles indicate the radii of the stellar distributions. The shaded ellipse marks the spatial distribution found from GMM analysis. \emph{Bottom left:} Distance distribution from fitting a Gaussian kernel. The two vertical
dotted lines indicate the distance range adopted for
the cluster stars. \emph{Bottom right:} CMD of the cluster stars. The solid line is the best fit isochrone.
}
\label{fig:4panel}
\end{figure}

We iterated our procedure until all six parameters -- center and radius of proper motion distribution, center and radius of spatial distribution, distance, and angle -- converged. In most cases, convergence was achieved when setting the power index $p$ in the weight function
to $p=0.5$ for cluster radii larger than $2^{\circ}$, and to $p=0.3$ for smaller radii.
In some cases, we varied the $p$ values to achieve convergence.

\begin{SCfigure*}[1][!t]
\centering
\includegraphics[width=0.775\textwidth]{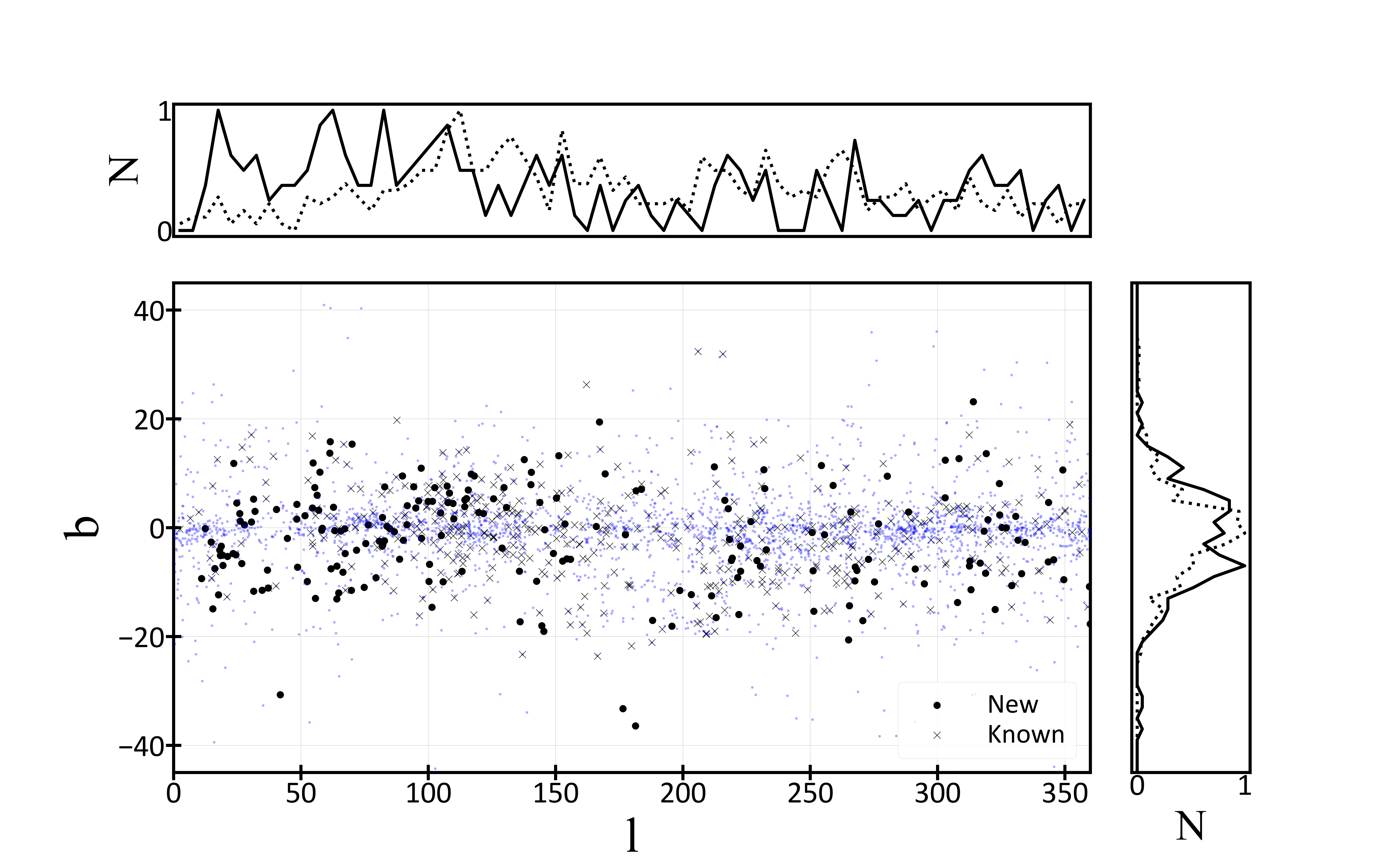}
\caption{Distribution of open clusters in Galactic coordinates $l,b$. Newly found clusters, known clusters, and MWSC clusters  are plotted as black dots, crosses, and small blue dots, respectively. The frequency distributions of visually identified new and known clusters are shown along the $l$ and $b$ axes, with $N$ being the number; solid lines represent the newly found clusters, dotted lines indicate the known clusters.}
\label{fig:l_v}
\end{SCfigure*}

\subsection{Color--Magnitude Diagrams}
	
To probe if they actually are physical star clusters, we constructed color--magnitude diagrams for our candidate clusters. We compared each CMD to the metallicity $Z = 0.02$  isochrones of the PARSEC library \citep{bre12}, updated for the \emph{Gaia} DR2 passbands\footnote{\url{http://stev.oapd.inaf.it/cgi-bin/cmd}} with the photometric calibration of \citet{eva18}.
We used least-squares fitting of isochrones to observed cluster CMDs. The CMDs were corrected for interstellar reddening $E(G_{\rm{BP}}-G_{\rm{RP}})$ and the extinction in the G band, $A_{\rm{G}}$, using the extinction values from \emph{Gaia} DR2. For stars with no extinction estimates in \emph{Gaia} DR2, we used the average of the extinction values of the cluster stars. We used the most probable parallax (mode of distribution) as cluster distance.  We found that all candidate clusters except of two have CMDs that are well described by theoretical isochrones; we selected the best-fitted isochrones to estimate the cluster ages. Thus, among the 655 visually selected cluster candidates, 653 are physical open
clusters ($99.7\%$).  The two cluster candidates that failed the validation, UPK~034 and UPK~036, present peculiar CMDs. UPK~034 has a main sequence that covers only $\sim3 mag$ with a large color spread, and UPK~036 has a main sequence too broad to be considered a cluster main sequence. Both are extremely young ($\sim6$ Myr) if they are actual open clusters.
Some of our clusters have small numbers of member stars ($N\lesssim50$)
and could be stellar associations or small stellar aggregations, but we nevertheless refer to them as ``open clusters''.

\subsection{Cross-Matching with Other Catalogs}
	
  There are a number of catalogs of open clusters which can be used to check whether the open clusters identified in the present study are new findings or not. We selected the three catalogs DALM, MWSC, and COCD for cross-matching of the 655 visually selected clusters. The total number of clusters listed in the three catalogs is 5755; due to duplications, the number of clusters to be cross-matched with our findings is $\sim$3000. In addition, we cross-matched our sample with the 1229 clusters from DALM by applying UPMASK to the \emph{Gaia} DR2  \citep{can18} to check the completeness of our search method. We also cross--matching our clusters with those from recent publications \citep{ros16, cas18, can19}.
  
We considered a candidate cluster to be a known cluster if it has coordinates, proper motions, and a distance that satisfy the following conditions. Firstly, the candidate cluster should be located within the cluster angular radius from the known clusters. That is, $\Delta\theta < r_{cl}$ where $\Delta\theta$ is the angular distance between the candidate cluster and a known cluster and $r_{cl}$ is the angular radius
of the candidate cluster measured in $l-b$ space. We used $r_{\rm{GMM}}$ for $r_{cl}$. Secondly, the clusters have similar proper motions, meaning here
\[
\left[(\mu_{\alpha}\cos\delta - \mu_{\alpha_{C}}\cos\delta_{C})^{2} + (\mu_{\delta} - \mu_{\delta_{C}})^{2}\right]^{1/2} < r_{cl(pm)}
\]
where $\mu_{\alpha}\cos\delta$ and $\mu_{\delta}$  are the proper motions of the candidate cluster and $\mu_{\alpha_{C}}\cos\delta_{C}$ and $\mu_{\delta_{C}}$ are those of the known cluster, respectively; $r_{cl(pm)}$ is the radius of the candidate cluster in proper motion space. Thirdly, $d_{min} < d_{C} < d_{max}$ where $d_{min}$ and $d_{max}$ are the minimum and maximum distance allowed for the candidate cluster and $d_{C}$ is the distance of the known cluster.

We cross-matched all 655 visually selected cluster candidates with known open clusters as described above.  We assigned numeric codes from 0 to 3 for candidate clusters to quantify the degree of agreement. If a candidate cluster did not satisfy any of our three conditions, we assigned code 0. We assigned code 1 to candidate clusters that satisfied only the first condition. Code 2 was assigned to candidate clusters which satisfy the first condition along with either the second or the third condition. If the candidate cluster satisfied all the three conditions, we assigned code 3. Since we wanted to restrict our sample to genuinely unknown clusters, we considered all clusters with a code other than 0 as known. When applying this threshold, we found 207 new open clusters among the 653 clusters that passed the CMD validation test.

\section{Results}
\label{sec:results}
\subsection{Catalog}

We summarize all 655 cluster candidates selected from visual inspection of \emph{Gaia} DR2 in an online catalog,\footnote{\url{https://sites.google.com/ushs.hs.kr/upk}} including the two cluster candidates that failed the CMD validation. We provide the cluster data along with figures visualizing the cluster parameters. The catalog provides coordinates, proper motions, and shape parameters derived from GMM analysis. Each figure presents the distribution of cluster stars in proper motion space and $l-b$ space, the distance distribution, and the CMD. An example is given in Figure~\ref{fig:4panel}. Distance distributions include all stars identified as cluster members in proper motion space and in $l-b$ space.

\begin{figure}[!t]
	\centering
	\includegraphics[width=1\columnwidth]{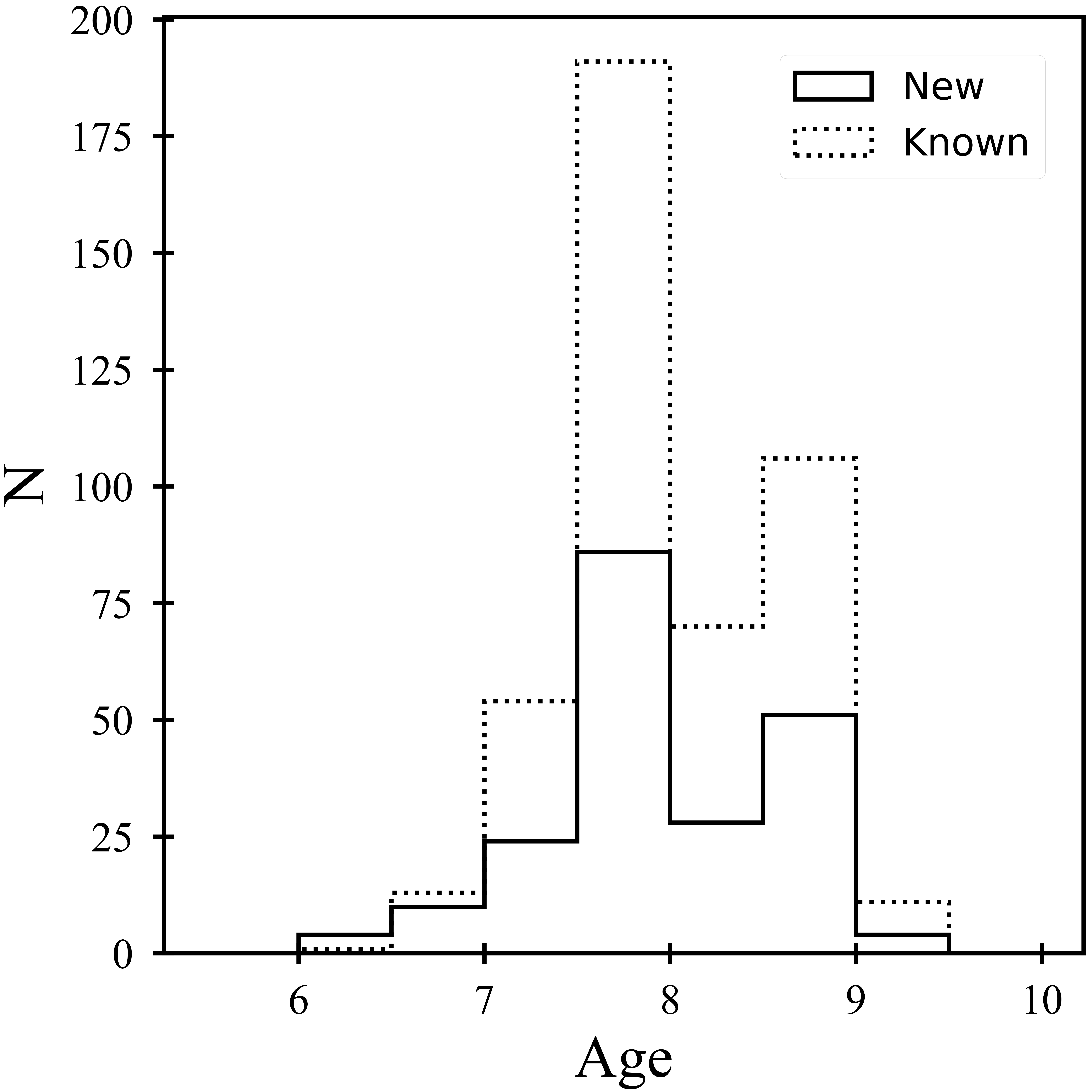}
	\caption{Frequency distribution of cluster ages. Solid lines represent new clusters, dotted lines indicate known clusters. }
	\label{fig:Hage}
\end{figure}

We present the basic parameters of the newly discovered 207 open clusters
in Table~\ref{tab2}. We give Galactic coordinates, proper motions,  parallaxes,
$r_{\rm{GMM}}$, $r_{\rm{c}}$, numbers of stars within
$r_{\rm{GMM}}$ ($N$), and cluster ages from the best-fit $Z=0.02$ PARSEC isochrones.
Other parameters such as the magnitude and colors of the
brightest cluster star, shape parameters (semi-major and minor axis,
ellipticity and position angle $\theta$) are given in the online catalog.
The radius $r_{\rm{GMM}}$ is not the conventional cluster radius but it
roughly defines the regions where cluster stars, including escaping stars,
are likely to be found.

\begin{table*}[tp]
\centering 
\caption{Physical parameters of newly discovered open clusters.}
\setlength{\tabcolsep}{3.75pt}
\label{tab2}
\begin{tabular}{lccccccccc}
\toprule 
	ID &
	$l$~($^\circ$)  &  $b$~($^\circ$) &
	$\mu_{\alpha^*}$~(mas yr$^{-1}$)  &
	$\mu_{\delta}$~(mas yr$^{-1}$) &
	$d$~(pc) & $r_{\rm{GMM}}~(^\prime)^{\rm a}$ &
	$r_{\rm{c}}~(^\prime)^{\rm b}$ & $N^{\rm c}$ &
	$\rm\log(age)$ \\ 
\midrule 
UPK~4* & 10.96 & -9.35 & -2.03 (0.09) & -6.40 (0.11) & 784 (24) & 22.48 & 11.25 (0.07) & 37 (89.66) & 8.45 \\
UPK~5 & 12.39 & -0.16 & 0.74 (0.10) & -8.38 (0.17) & 550 (16) & 23.30 & 8.26 (0.12) & 25 (34.69) & 8.00 \\
UPK~7 & 14.74 & -2.64 & -0.28 (0.14) & -5.87 (0.08) & 769 (22) & 12.12 & 5.99 (0.08) & 19 (51.88) & 8.40 \\
UPK~12 & 15.37 & -14.92 & 0.58 (0.10) & -6.61 (0.12) & 644 (29) & 15.82 & 10.89 (0.02) & 24 (39.24) & 7.90 \\
UPK~13 & 16.18 & -7.51 & -1.15 (0.10) & -4.12 (0.10) & 915 (25) & 12.11 & 2.67 (0.02) & 45 (85.00) & 7.90 \\
UPK~16 & 17.63 & -12.34 & -0.40 (0.18) & -5.14 (0.19) & 802 (39) & 23.81 & 6.82 (0.08) & 29 (55.16) & 7.90 \\
UPK~17 & 18.11 & -4.11 & 0.90 (0.11) & -7.92 (0.12) & 659 (20) & 27.44 & 8.55 (0.08) & 39 (83.82) & 8.90 \\
UPK~18 & 18.34 & -5.11 & 3.08 (0.14) & 0.87 (0.17) & 753 (37) & 14.48 & 6.54 (0.03) & 24 (45.65) & 8.45 \\
UPK~19 & 18.70 & -3.39 & -1.81 (0.28) & -9.33 (0.29) & 289 (27) & 94.13 & 44.34 (0.23) & 119 (119.00) & 8.50 \\
UPK~20 & 19.24 & -5.05 & 3.32 (0.15) & 1.14 (0.14) & 472 (23) & 44.32 & 22.11 (0.09) & 40 (65.44) & 8.80 \\
UPK~21 & 19.38 & -6.93 & 7.82 (0.10) & 0.73 (0.10) & 587 (13) & 20.98 & 9.92 (0.03) & 22 (53.55) & 8.95 \\
UPK~23 & 21.14 & -5.27 & -0.90 (0.12) & -2.08 (0.13) & 952 (36) & 13.04 & 6.38 (0.14) & 19 (52.90) & 8.55 \\
UPK~24 & 23.33 & -4.74 & -3.75 (0.12) & -9.79 (0.15) & 498 (18) & 72.33 & 18.24 (0.97) & 32 (44.49) & 8.90 \\
UPK~25 & 23.59 & 11.81 & -0.31 (0.09) & 0.19 (0.12) & 629 (31) & 50.21 & 24.68 (0.43) & 19 (46.17) & 8.95 \\
UPK~26 & 24.44 & -5.00 & 0.08 (0.10) & -4.42 (0.08) & 875 (26) & 26.23 & 15.27 (0.16) & 42 (78.92) & 8.05 \\
UPK~27 & 24.78 & 4.54 & 2.36 (0.10) & -2.14 (0.14) & 899 (29) & 14.16 & 5.86 (0.03) & 36 (116.70) & 9.35 \\
UPK~28 & 25.98 & 2.60 & 0.81 (0.12) & -2.49 (0.13) & 882 (33) & 13.44 & 0.98 (0.25) & 28 (90.77) & 8.90 \\
UPK~29 & 26.13 & 1.19 & 0.45 (0.12) & 2.83 (0.11) & 656 (13) & 13.52 & 13.38 (0.12) & 22 (88.07) & 8.75 \\
UPK~31 & 26.75 & -6.59 & -0.32 (0.09) & 1.21 (0.08) & 540 (13) & 18.55 & 10.70 (0.09) & 17 (36.62) & 7.55 \\
UPK~33 & 27.78 & 0.52 & -0.13 (0.15) & -10.95 (0.17) & 508 (11) & 14.29 & 12.30 (0.11) & 16 (33.94) & 8.90 \\
UPK~38 & 30.54 & 1.04 & -1.11 (0.11) & -5.27 (0.10) & 572 (7) & 8.75 & 1.52 (0.11) & 8 (11.53) & 6.45 \\
UPK~39 & 31.37 & 5.24 & 3.29 (0.39) & -8.65 (0.41) & 431 (27) & 29.77 & 13.52 (0.07) & 62 (117.00) & 6.45 \\
UPK~40 & 31.44 & -11.69 & 0.29 (0.15) & -9.13 (0.17) & 641 (31) & 58.58 & 38.19 (0.22) & 32 (60.39) & 8.75 \\
UPK~41 & 31.95 & 3.00 & 2.51 (0.22) & -8.13 (0.23) & 480 (19) & 43.23 & 34.34 (0.16) & 61 (115.35) & 6.85 \\
UPK~42 & 34.73 & -11.50 & -0.46 (0.12) & -5.03 (0.11) & 592 (15) & 22.54 & 5.99 (0.03) & 41 (77.45) & 8.75 \\
UPK~45 & 36.84 & -7.79 & -0.62 (0.11) & -2.16 (0.10) & 862 (51) & 110.95 & 27.29 (0.24) & 179 (335.37) & 7.90 \\
UPK~46 & 37.13 & -11.08 & 0.11 (0.15) & -8.60 (0.17) & 680 (25) & 27.27 & 17.18 (0.27) & 23 (43.40) & 7.90 \\
UPK~50 & 40.40 & 3.35 & -1.33 (0.11) & -3.79 (0.14) & 995 (34) & 20.49 & 14.14 (0.14) & 51 (123.74) & 8.70 \\
UPK~51 & 41.87 & -30.70 & 5.06 (0.16) & 2.07 (0.11) & 567 (22) & 34.35 & 18.46 (0.10) & 21 (24.86) & 8.00 \\
UPK~52 & 44.69 & -1.96 & -0.16 (0.05) & -5.97 (0.10) & 818 (28) & 12.14 & 4.24 (0.05) & 9 (24.97) & 8.05 \\
UPK~53 & 48.32 & 1.56 & 2.44 (0.12) & 1.08 (0.09) & 591 (13) & 26.97 & 14.93 (0.05) & 43 (81.31) & 8.90 \\
UPK~54 & 48.37 & 4.28 & -0.57 (0.12) & -5.88 (0.12) & 827 (31) & 26.16 & 14.53 (0.11) & 60 (128.62) & 7.90 \\
UPK~55 & 48.66 & -7.27 & -1.43 (0.11) & -6.86 (0.09) & 777 (28) & 20.40 & 7.93 (0.05) & 30 (48.86) & 7.95 \\
UPK~56 & 51.57 & 2.20 & -0.07 (0.16) & -6.74 (0.15) & 760 (27) & 63.41 & 34.18 (0.17) & 87 (186.72) & 8.75 \\
UPK~57 & 52.49 & -9.88 & -0.45 (0.08) & -3.56 (0.08) & 819 (36) & 26.31 & 15.71 (0.12) & 14 (22.90) & 8.75 \\
UPK~62 & 54.52 & 3.61 & -0.43 (0.12) & -5.41 (0.08) & 895 (40) & 16.80 & 3.41 (0.07) & 26 (104.80) & 6.75 \\
UPK~64 & 54.76 & 11.90 & -0.93 (0.11) & -5.03 (0.12) & 442 (57) & 84.51 & 30.75 (0.40) & 21 (24.89) & 8.00 \\
UPK~65 & 55.40 & 7.39 & -1.86 (0.08) & -3.56 (0.08) & 758 (20) & 22.06 & 10.40 (0.04) & 23 (43.44) & 8.90 \\
UPK~66 & 55.69 & -12.98 & 0.51 (0.08) & -9.93 (0.10) & 669 (13) & 21.99 & 16.40 (0.18) & 16 (33.98) & 8.10 \\
UPK~70 & 56.33 & 5.96 & 0.33 (0.21) & -5.38 (0.17) & 493 (28) & 28.30 & 24.80 (0.13) & 27 (37.54) & 7.10 \\
UPK~71 & 56.93 & 3.20 & 1.30 (0.28) & -5.48 (0.25) & 524 (37) & 16.02 & 5.50 (0.03) & 40 (85.27) & 6.85 \\
UPK~72 & 57.42 & 10.19 & -0.68 (0.20) & -4.41 (0.21) & 454 (43) & 75.81 & 95.39 (1.27) & 57 (67.48) & 7.80 \\
UPK~73 & 58.13 & -0.47 & 0.74 (0.11) & -6.51 (0.13) & 634 (22) & 11.24 & 7.87 (0.09) & 9 (29.39) & 7.25 \\
UPK~74 & 58.41 & -0.11 & -0.11 (0.10) & -3.33 (0.10) & 571 (38) & 26.03 & 5.12 (0.02) & 16 (18.90) & 7.95 \\
UPK~77 & 61.33 & 13.69 & 1.08 (0.05) & -4.06 (0.07) & 353 (25) & 19.72 & 7.33 (0.25) & 7 (8.37) & 7.40 \\
UPK~78 & 61.49 & 15.81 & 0.31 (0.28) & -3.52 (0.22) & 375 (33) & 92.39 & 133.03 (2.22) & 50 (51.13) & 7.75 \\
UPK~79 & 61.81 & -7.56 & -2.10 (0.14) & -5.47 (0.10) & 913 (37) & 27.85 & 12.35 (0.08) & 23 (55.80) & 7.80 \\
UPK~80 & 62.75 & 3.76 & 1.19 (0.10) & -1.01 (0.09) & 876 (49) & 35.63 & 7.09 (0.34) & 28 (53.04) & 7.90 \\
UPK~82 & 63.22 & -0.54 & 2.34 (0.11) & -2.13 (0.10) & 537 (21) & 37.87 & 25.39 (0.09) & 41 (48.48) & 7.90 \\
UPK~84 & 64.07 & -13.10 & -6.25 (0.07) & -9.25 (0.09) & 904 (25) & 24.29 & 13.72 (0.06) & 38 (81.56) & 8.75 \\
UPK~85 & 64.12 & -7.06 & 2.12 (0.10) & -1.61 (0.07) & 887 (26) & 30.12 & 28.72 (1.15) & 11 (24.06) & 7.60 \\
UPK~88 & 64.83 & -11.99 & 10.29 (0.12) & -5.35 (0.19) & 293 (19) & 64.38 & 26.84 (0.41) & 16 (16.00) & 7.80 \\
UPK~90 & 65.63 & -0.63 & -0.83 (0.10) & -5.72 (0.08) & 793 (32) & 13.21 & 10.56 (0.23) & 24 (45.46) & 7.65 \\
UPK~91 & 66.48 & -8.18 & 2.04 (0.13) & -3.97 (0.09) & 764 (46) & 76.16 & 31.67 (0.23) & 30 (48.98) & 8.00 \\
UPK~93 & 67.19 & -0.20 & -0.85 (0.10) & -8.28 (0.11) & 685 (34) & 32.51 & 7.51 (0.08) & 33 (53.88) & 8.60 \\
\bottomrule
\end{tabular}
\tabnote{
*UPK stands for the names of the authors' institutions: Ulsan Science High School, Pusan National University, Korea Astronomy and Space Science Institute. Numbers in  parenthesis are errors. \\ 
$^{\rm a}$GMM cluster radius. \\ 
$^{\rm b}$King profile core radius. \\ 
$^{\rm c}$Number of stars within the best-fit ellipse. Numbers in parenthesis are numbers of member stars corrected for the incompleteness due to the magnitude cut at $G=18$.
}
\end{table*}

\addtocounter{table}{-1}
\begin{table*}[tp]
\centering 
\caption{\emph{Continued}}
\setlength{\tabcolsep}{3.75pt}
\begin{tabular}{lccccccccc}
\toprule 
	ID &
	$l$~($^\circ$)  &  $b$~($^\circ$) &
	$\mu_{\alpha^*}$~(mas yr$^{-1}$)  &
	$\mu_{\delta}$~(mas yr$^{-1}$) &
	$d$~(pc) & $r_{\rm{GMM}}~(^\prime)^{\rm a}$ &
	$r_{\rm{c}}~(^\prime)^{\rm b}$ & $N^{\rm c}$ &
	$\rm\log(age)$ \\ 
\midrule 
UPK~94 & 67.35 & -4.74 & 2.12 (0.06) & 1.06 (0.06) & 941 (32) & 23.49 & 9.14 (0.09) & 20 (56.01) & 7.90 \\
UPK~99 & 69.80 & -11.52 & 0.27 (0.17) & -7.84 (0.21) & 783 (37) & 39.05 & 19.86 (0.74) & 20 (32.86) & 8.10 \\
UPK~101 & 70.08 & 15.36 & 1.44 (0.33) & -1.63 (0.34) & 363 (42) & 100.06 & 112.27 (1.11) & 103 (105.30) & 7.65 \\
UPK~104 & 71.82 & -4.14 & 3.67 (0.17) & 1.53 (0.17) & 836 (35) & 40.53 & 28.16 (0.33) & 34 (64.22) & 7.95 \\
UPK~108 & 74.84 & -10.97 & 1.02 (0.11) & -4.72 (0.09) & 840 (26) & 52.71 & 13.41 (0.21) & 28 (52.95) & 7.65 \\
UPK~109 & 75.43 & -2.91 & -2.37 (0.08) & -5.67 (0.10) & 725 (23) & 17.61 & 0.60 (0.44) & 12 (22.67) & 7.65 \\
UPK~110 & 76.38 & 0.50 & 2.90 (0.11) & -1.17 (0.08) & 857 (15) & 30.93 & 15.10 (0.12) & 11 (23.61) & 8.65 \\
UPK~113 & 79.43 & -9.19 & 2.16 (0.08) & -2.34 (0.13) & 679 (10) & 22.91 & 8.98 (0.13) & 15 (24.51) & 7.85 \\
UPK~116 & 80.78 & -2.48 & 3.28 (0.08) & -2.94 (0.06) & 435 (19) & 31.69 & 11.21 (0.06) & 14 (14.32) & 7.75 \\
UPK~118 & 81.92 & -3.39 & -1.65 (0.08) & -3.50 (0.09) & 940 (40) & 73.88 & 19.62 (0.10) & 38 (92.76) & 8.90 \\
UPK~119 & 81.99 & 1.87 & -0.65 (0.06) & -2.00 (0.07) & 769 (11) & 16.50 & 11.93 (1.30) & 18 (43.67) & 8.75 \\
UPK~120 & 82.12 & -3.14 & 2.76 (0.16) & -1.44 (0.15) & 666 (20) & 38.40 & 6.37 (0.09) & 22 (47.22) & 7.75 \\
UPK~121 & 82.77 & -2.43 & 0.57 (0.15) & -1.87 (0.12) & 951 (52) & 48.17 & 27.03 (0.18) & 43 (139.39) & 8.95 \\
UPK~122 & 82.86 & 7.50 & -1.34 (0.07) & -9.26 (0.04) & 900 (25) & 36.29 & 51.93 (5.82) & 13 (34.41) & 9.40 \\
UPK~126 & 83.78 & 0.27 & -1.46 (0.16) & -1.89 (0.18) & 823 (36) & 9.61 & 1.90 (0.06) & 15 (41.32) & 6.75 \\
UPK~127 & 84.82 & -0.17 & -1.29 (0.39) & -3.08 (0.42) & 796 (48) & 26.37 & 12.74 (0.03) & 161 (439.08) & 6.25 \\
UPK~131 & 86.66 & -0.73 & 2.83 (0.10) & -1.09 (0.07) & 976 (25) & 19.63 & 8.44 (0.05) & 24 (77.61) & 8.75 \\
UPK~136 & 88.73 & -5.79 & -0.03 (0.16) & -0.60 (0.12) & 643 (35) & 50.28 & 35.17 (0.35) & 42 (58.40) & 7.85 \\
UPK~137 & 89.84 & 9.50 & -2.34 (0.10) & -6.04 (0.09) & 838 (28) & 22.02 & 9.22 (0.02) & 19 (46.25) & 7.50 \\
UPK~138 & 90.21 & -2.32 & -0.41 (0.15) & -1.78 (0.13) & 681 (12) & 10.73 & 1.63 (0.18) & 11 (35.66) & 6.50 \\
UPK~143 & 91.62 & 0.55 & 1.07 (0.12) & -2.46 (0.11) & 861 (26) & 50.42 & 9.34 (0.07) & 65 (207.93) & 8.85 \\
UPK~144 & 91.74 & 4.04 & 1.17 (0.38) & -3.41 (0.26) & 557 (29) & 20.50 & 6.17 (0.05) & 27 (66.34) & 6.45 \\
UPK~147 & 94.25 & 7.51 & -2.39 (0.19) & -4.83 (0.23) & 883 (35) & 106.71 & 98.88 (0.70) & 188 (448.76) & 7.25 \\
UPK~150 & 95.06 & 3.62 & -4.25 (0.09) & -3.99 (0.09) & 980 (33) & 48.73 & 15.75 (0.06) & 40 (129.11) & 7.20 \\
UPK~152 & 96.18 & 4.94 & 3.73 (0.10) & -1.17 (0.12) & 739 (35) & 47.42 & 13.29 (0.20) & 19 (46.38) & 9.00 \\
UPK~155 & 97.22 & 10.93 & -2.24 (0.09) & -3.86 (0.15) & 887 (23) & 45.48 & 10.40 (0.13) & 23 (56.15) & 7.55 \\
UPK~156 & 97.32 & -1.95 & -0.32 (0.10) & -1.91 (0.08) & 902 (27) & 26.36 & 10.72 (0.15) & 14 (26.31) & 7.90 \\
UPK~160 & 98.40 & 15.02 & 3.68 (0.21) & 0.29 (0.18) & 480 (15) & 29.30 & 2.03 (0.20) & 14 (22.89) & 6.70 \\
UPK~164 & 99.92 & 4.82 & -1.46 (0.18) & -3.26 (0.17) & 933 (45) & 71.26 & 16.23 (0.16) & 124 (339.67) & 7.25 \\
UPK~166 & 100.27 & -9.88 & -0.92 (0.12) & -3.26 (0.17) & 651 (27) & 75.77 & 17.49 (0.12) & 131 (181.84) & 7.60 \\
UPK~167 & 100.48 & -6.74 & 1.56 (0.17) & -0.09 (0.16) & 545 (33) & 101.05 & 38.98 (0.17) & 89 (105.17) & 8.00 \\
UPK~168 & 101.43 & -14.64 & -0.53 (0.13) & -3.95 (0.13) & 596 (21) & 72.94 & 16.40 (0.13) & 99 (137.16) & 7.65 \\
UPK~169 & 101.63 & 4.84 & -1.95 (0.15) & -2.36 (0.16) & 842 (30) & 52.97 & 10.24 (0.10) & 64 (152.85) & 7.10 \\
UPK~172 & 102.55 & 7.35 & -2.06 (0.12) & -3.04 (0.10) & 909 (34) & 17.14 & 3.74 (0.04) & 35 (84.81) & 7.15 \\
UPK~178 & 104.86 & 2.71 & -4.34 (0.15) & -3.36 (0.14) & 949 (21) & 5.73 & 1.53 (0.03) & 15 (48.42) & 8.50 \\
UPK~180 & 105.11 & -1.44 & -0.61 (0.17) & 0.21 (0.16) & 832 (38) & 45.13 & 18.83 (0.08) & 40 (75.64) & 8.55 \\
UPK~185 & 105.82 & -9.95 & 2.66 (0.16) & -1.93 (0.17) & 552 (23) & 31.02 & 5.25 (0.03) & 83 (98.27) & 7.95 \\
UPK~189 & 107.39 & 7.64 & -2.65 (0.10) & -3.32 (0.11) & 1001 (40) & 44.04 & 44.48 (0.41) & 31 (86.50) & 7.20 \\
UPK~191 & 107.68 & 4.61 & -1.04 (0.16) & -1.87 (0.17) & 874 (37) & 37.11 & 0.02 (0.00) & 34 (107.73) & 6.90 \\
UPK~194 & 108.29 & 6.36 & -0.76 (0.31) & -3.09 (0.33) & 924 (33) & 25.37 & 8.27 (0.02) & 92 (250.28) & 7.35 \\
UPK~198 & 109.68 & 4.45 & -4.90 (0.10) & -2.14 (0.11) & 844 (26) & 50.72 & 6.00 (0.07) & 27 (74.92) & 8.60 \\
UPK~201 & 109.99 & 1.64 & -3.95 (0.16) & -3.37 (0.09) & 919 (15) & 16.20 & 1.80 (0.22) & 18 (58.21) & 6.60 \\
UPK~214 & 113.24 & -8.02 & 0.79 (0.21) & -0.72 (0.21) & 810 (46) & 95.14 & 53.33 (0.23) & 111 (180.78) & 8.75 \\
UPK~219 & 114.27 & 3.86 & -1.72 (0.06) & -2.48 (0.07) & 801 (19) & 23.39 & 6.11 (0.11) & 15 (32.24) & 8.90 \\
UPK~220 & 114.35 & 5.10 & -2.41 (0.13) & -2.59 (0.12) & 967 (28) & 23.57 & 6.15 (0.04) & 102 (326.29) & 8.75 \\
UPK~224 & 114.99 & 5.35 & -1.41 (0.11) & -0.10 (0.12) & 779 (21) & 20.33 & 4.60 (0.09) & 30 (97.25) & 6.85 \\
UPK~226 & 115.70 & 6.93 & -1.55 (0.09) & -3.13 (0.07) & 861 (14) & 26.66 & 13.95 (0.13) & 25 (60.65) & 8.80 \\
UPK~230 & 116.88 & 9.85 & 4.90 (0.10) & -0.79 (0.12) & 535 (17) & 31.45 & 24.66 (0.37) & 19 (26.29) & 8.35 \\
UPK~233 & 118.06 & 9.50 & 6.89 (0.18) & 0.69 (0.19) & 493 (13) & 48.50 & 22.95 (0.33) & 18 (25.00) & 7.90 \\
UPK~237 & 119.80 & 2.81 & -0.59 (0.15) & -3.75 (0.17) & 735 (21) & 25.59 & 5.26 (0.04) & 53 (112.56) & 7.65 \\
UPK~241 & 121.62 & 2.60 & 0.17 (0.13) & -0.20 (0.10) & 949 (25) & 29.81 & 13.53 (0.05) & 27 (87.31) & 8.90 \\
UPK~252 & 125.62 & 5.31 & -1.46 (0.14) & -0.85 (0.16) & 858 (47) & 103.73 & 20.87 (0.18) & 47 (130.41) & 7.40 \\
UPK~260 & 128.99 & -3.75 & 2.66 (0.12) & -1.86 (0.14) & 825 (42) & 75.73 & 31.92 (0.36) & 54 (101.83) & 7.90 \\
UPK~262 & 129.73 & 7.38 & -1.97 (0.07) & 0.77 (0.08) & 825 (32) & 52.06 & 44.28 (0.29) & 14 (38.57) & 8.50 \\
UPK~265 & 130.67 & 3.73 & -0.51 (0.09) & -0.37 (0.11) & 864 (32) & 35.01 & 12.74 (0.08) & 40 (96.93) & 8.60 \\
\bottomrule
\end{tabular}
\tabnote{
*UPK stands for the names of the authors' institutions: Ulsan Science High School, Pusan National University, Korea Astronomy and Space Science Institute. Numbers in  parenthesis are errors. \\ 
$^{\rm a}$GMM cluster radius. \\ 
$^{\rm b}$King profile core radius. \\ 
$^{\rm c}$Number of stars within the best-fit ellipse. Numbers in parenthesis are numbers of member stars corrected for the incompleteness due to the magnitude cut at $G=18$.
}
\end{table*}

\addtocounter{table}{-1}
\begin{table*}[tp]
\centering 
\caption{\emph{Continued}}
\setlength{\tabcolsep}{3.75pt}
\begin{tabular}{lccccccccc}
\toprule 
	ID &
	$l$~($^\circ$)  &  $b$~($^\circ$) &
	$\mu_{\alpha^*}$~(mas yr$^{-1}$)  &
	$\mu_{\delta}$~(mas yr$^{-1}$) &
	$d$~(pc) & $r_{\rm{GMM}}~(^\prime)^{\rm a}$ &
	$r_{\rm{c}}~(^\prime)^{\rm b}$ & $N^{\rm c}$ &
	$\rm\log(age)$ \\ 
\midrule 
UPK~281 & 135.79 & -8.00 & -4.07 (0.16) & 0.39 (0.13) & 624 (59) & 107.05 & 41.18 (0.50) & 38 (52.77) & 8.10 \\
UPK~282 & 136.05 & -17.30 & -2.79 (0.12) & -3.19 (0.12) & 807 (40) & 19.62 & 7.71 (0.05) & 22 (42.17) & 7.90 \\
UPK~287 & 137.67 & 12.52 & -2.38 (0.27) & -3.46 (0.26) & 630 (75) & 117.96 & 65.11 (0.76) & 52 (97.21) & 8.45 \\
UPK~292 & 140.30 & 7.91 & 0.23 (0.17) & -8.12 (0.19) & 444 (33) & 66.80 & 9.92 (0.17) & 28 (45.81) & 8.80 \\
UPK~294 & 140.45 & 10.18 & -1.27 (0.13) & 2.25 (0.18) & 746 (17) & 50.23 & 59.44 (0.86) & 27 (51.05) & 8.95 \\
UPK~296 & 142.50 & -9.83 & 1.97 (0.27) & -6.75 (0.31) & 538 (49) & 66.41 & 23.15 (0.11) & 49 (57.94) & 7.95 \\
UPK~300 & 143.78 & 4.64 & 0.25 (0.13) & -0.75 (0.10) & 962 (42) & 35.87 & 15.55 (0.08) & 21 (57.85) & 8.60 \\
UPK~303 & 144.56 & -18.03 & 18.25 (0.38) & -24.33 (0.49) & 212 (8) & 149.18 & 32.94 (0.68) & 33 (33.00) & 7.95 \\
UPK~305 & 145.35 & -19.04 & 2.96 (0.13) & -6.50 (0.10) & 411 (31) & 92.07 & 25.93 (0.43) & 32 (32.72) & 7.95 \\
UPK~307 & 145.71 & -0.37 & -2.13 (0.18) & -1.82 (0.17) & 893 (41) & 71.62 & 18.20 (0.16) & 54 (128.96) & 7.25 \\
UPK~310 & 149.14 & -4.72 & -2.66 (0.07) & 0.61 (0.11) & 620 (62) & 79.73 & 0.06 (0.00) & 14 (19.43) & 7.75 \\
UPK~312 & 150.30 & 5.44 & 0.20 (0.07) & -0.73 (0.07) & 702 (22) & 49.58 & 0.02 (0.00) & 23 (44.09) & 8.90 \\
UPK~317 & 151.23 & 13.22 & -1.11 (0.19) & -1.46 (0.17) & 900 (52) & 77.98 & 12.95 (0.43) & 42 (89.94) & 7.60 \\
UPK~322 & 152.74 & -6.13 & -1.39 (0.15) & -2.99 (0.15) & 955 (26) & 26.83 & 9.49 (0.05) & 60 (127.43) & 7.55 \\
UPK~325 & 153.66 & 0.68 & -1.55 (0.08) & -2.30 (0.08) & 762 (28) & 17.30 & 17.35 (0.15) & 11 (38.59) & 7.95 \\
UPK~326 & 154.42 & -5.74 & -2.16 (0.30) & 0.10 (0.34) & 646 (50) & 77.08 & 49.70 (0.42) & 40 (65.39) & 8.95 \\
UPK~333 & 155.85 & -5.81 & 2.53 (0.14) & -6.47 (0.13) & 723 (41) & 36.65 & 12.76 (0.08) & 30 (56.61) & 8.80 \\
UPK~347 & 165.93 & 0.22 & -0.46 (0.12) & -3.70 (0.06) & 917 (30) & 35.90 & 3.93 (0.25) & 19 (35.89) & 7.90 \\
UPK~350 & 167.21 & 19.42 & -4.88 (0.17) & -6.39 (0.17) & 444 (23) & 88.17 & 30.22 (0.13) & 37 (37.83) & 8.00 \\
UPK~357 & 169.46 & 9.90 & -0.33 (0.12) & -1.21 (0.11) & 716 (59) & 162.52 & 32.63 (0.36) & 36 (58.51) & 8.20 \\
UPK~367 & 176.45 & -33.27 & 24.93 (0.67) & -24.34 (0.52) & 162 (46) & 294.39 & 123.78 (0.65) & 42 (42.00) & 7.95 \\
UPK~369 & 177.42 & -1.28 & 0.08 (0.06) & -2.89 (0.08) & 731 (25) & 43.64 & 14.43 (0.14) & 18 (38.69) & 8.85 \\
UPK~378 & 181.37 & -36.43 & 21.48 (0.31) & -13.79 (0.33) & 182 (6) & 71.27 & 10.56 (0.29) & 13 (13.00) & 7.90 \\
UPK~379 & 181.60 & 6.79 & -0.97 (0.10) & -4.29 (0.08) & 765 (38) & 54.68 & 10.06 (0.11) & 22 (41.52) & 8.80 \\
UPK~381 & 183.70 & 7.05 & 0.48 (0.08) & -2.96 (0.08) & 708 (24) & 53.72 & 17.35 (0.06) & 28 (53.04) & 8.00 \\
UPK~385 & 188.05 & -17.04 & 1.87 (0.34) & -3.52 (0.28) & 320 (34) & 46.18 & 10.05 (0.15) & 38 (38.87) & 7.00 \\
UPK~394 & 195.66 & -18.11 & -0.52 (0.10) & -2.03 (0.10) & 802 (24) & 25.53 & 7.95 (0.06) & 27 (51.05) & 7.90 \\
UPK~398 & 198.75 & -11.55 & -3.34 (0.18) & -1.76 (0.18) & 452 (32) & 39.66 & 9.45 (0.06) & 40 (55.62) & 7.10 \\
UPK~402 & 203.26 & -12.30 & -1.15 (0.31) & -1.96 (0.28) & 429 (35) & 21.12 & 10.45 (0.04) & 43 (71.03) & 6.85 \\
UPK~416 & 211.24 & -12.52 & -1.26 (0.12) & 1.48 (0.15) & 871 (48) & 99.53 & 23.30 (0.25) & 34 (73.07) & 7.90 \\
UPK~418 & 212.31 & 11.18 & -2.68 (0.09) & -1.60 (0.10) & 828 (28) & 32.25 & 9.75 (0.08) & 17 (27.85) & 7.95 \\
UPK~422 & 213.03 & -16.53 & 0.25 (0.21) & -0.60 (0.22) & 304 (22) & 63.58 & 23.01 (0.08) & 71 (71.00) & 7.30 \\
UPK~429 & 216.44 & 5.03 & -2.01 (0.14) & 0.07 (0.12) & 865 (38) & 41.86 & 7.57 (0.12) & 76 (143.42) & 7.95 \\
UPK~431 & 217.79 & 3.49 & -2.44 (0.10) & 0.77 (0.11) & 729 (33) & 54.19 & 13.95 (0.12) & 55 (89.39) & 8.70 \\
UPK~433 & 218.28 & -2.15 & -0.78 (0.13) & -2.55 (0.11) & 780 (29) & 43.06 & 7.61 (0.23) & 51 (83.38) & 7.90 \\
UPK~436 & 219.00 & -6.02 & -4.10 (0.14) & 0.66 (0.17) & 842 (50) & 52.09 & 25.33 (0.13) & 67 (126.34) & 7.30 \\
UPK~438 & 219.35 & -5.60 & -6.59 (0.44) & 2.68 (0.37) & 449 (60) & 95.65 & 29.13 (0.16) & 52 (53.17) & 7.80 \\
UPK~442 & 221.51 & -9.17 & -2.54 (0.18) & -1.32 (0.17) & 640 (45) & 88.53 & 43.49 (0.32) & 55 (76.20) & 7.70 \\
UPK~445 & 221.93 & -15.95 & -2.41 (0.39) & 1.45 (0.33) & 667 (48) & 84.12 & 25.18 (0.24) & 131 (213.35) & 7.10 \\
UPK~447 & 222.57 & -3.38 & -3.60 (0.08) & 0.41 (0.11) & 962 (32) & 41.40 & 23.51 (0.06) & 30 (57.06) & 7.65 \\
UPK~448 & 222.64 & -8.00 & -2.61 (0.17) & 0.73 (0.17) & 900 (42) & 44.21 & 30.58 (0.76) & 37 (69.82) & 7.90 \\
UPK~452 & 226.65 & 1.13 & -5.80 (0.08) & 3.19 (0.08) & 623 (10) & 23.99 & 5.68 (0.09) & 9 (25.11) & 7.60 \\
UPK~456 & 229.07 & -6.03 & -7.62 (0.21) & 0.18 (0.17) & 506 (41) & 100.85 & 33.68 (0.10) & 31 (36.72) & 7.95 \\
UPK~457 & 230.41 & -7.05 & -3.81 (0.10) & 5.14 (0.10) & 949 (33) & 23.73 & 6.24 (0.08) & 16 (40.86) & 7.40 \\
UPK~467 & 231.78 & 10.65 & -6.99 (0.12) & -1.72 (0.11) & 579 (26) & 49.73 & 16.03 (0.09) & 49 (68.08) & 8.20 \\
UPK~468 & 232.13 & 7.20 & -1.58 (0.08) & 0.10 (0.11) & 871 (39) & 22.52 & 4.06 (0.22) & 22 (36.14) & 8.20 \\
UPK~470 & 232.73 & -4.05 & -4.81 (0.11) & 4.23 (0.10) & 999 (36) & 44.70 & 8.02 (0.10) & 55 (104.00) & 7.50 \\
UPK~492 & 250.73 & -0.88 & -7.43 (0.10) & 5.39 (0.07) & 713 (26) & 42.72 & 0.02 (0.00) & 17 (23.50) & 7.90 \\
UPK~494 & 251.06 & -7.92 & -5.35 (0.08) & 5.25 (0.10) & 881 (38) & 53.37 & 18.51 (0.16) & 19 (31.74) & 7.75 \\
UPK~495 & 251.37 & -15.37 & -2.42 (0.14) & 5.67 (0.17) & 664 (42) & 83.78 & 24.62 (0.31) & 36 (58.51) & 8.45 \\
UPK~499 & 254.36 & 11.42 & -6.75 (0.19) & 2.67 (0.17) & 1001 (70) & 110.31 & 113.82 (1.36) & 106 (199.17) & 7.90 \\
UPK~502 & 255.60 & -1.29 & -8.08 (0.17) & 5.61 (0.17) & 699 (19) & 15.96 & 3.76 (0.06) & 22 (35.99) & 7.85 \\
UPK~508 & 258.93 & 7.76 & -10.58 (0.09) & 6.25 (0.09) & 830 (20) & 26.33 & 2.98 (0.10) & 26 (63.08) & 8.60 \\
UPK~524 & 265.00 & -20.62 & -2.77 (0.19) & 8.81 (0.25) & 554 (30) & 77.75 & 38.60 (0.22) & 40 (47.29) & 8.00 \\
UPK~526 & 265.35 & -14.34 & -7.98 (0.19) & 7.42 (0.14) & 575 (23) & 86.20 & 27.79 (0.23) & 30 (41.63) & 7.30 \\

\bottomrule
\end{tabular}
\tabnote{
*UPK stands for the names of the authors' institutions: Ulsan Science High School, Pusan National University, Korea Astronomy and Space Science Institute. Numbers in  parenthesis are errors. \\ 
$^{\rm a}$GMM cluster radius. \\ 
$^{\rm b}$King profile core radius. \\ 
$^{\rm c}$Number of stars within the best-fit ellipse. Numbers in parenthesis are numbers of member stars corrected for the incompleteness due to the magnitude cut at $G=18$.
}
\end{table*}

\addtocounter{table}{-1}
\begin{table*}[t!]
\centering 
\caption{\emph{Continued}}
\setlength{\tabcolsep}{3.75pt}
\begin{tabular}{lccccccccc}
\toprule 
	ID &
	$l$~($^\circ$)  &  $b$~($^\circ$) &
	$\mu_{\alpha^*}$~(mas yr$^{-1}$)  &
	$\mu_{\delta}$~(mas yr$^{-1}$) &
	$d$~(pc) & $r_{\rm{GMM}}~(^\prime)^{\rm a}$ &
	$r_{\rm{c}}~(^\prime)^{\rm b}$ & $N^{\rm c}$ &
	$\rm\log(age)$ \\ 
\midrule 
UPK~528 & 265.90 & 2.90 & -7.69 (0.13) & 1.89 (0.13) & 953 (41) & 32.23 & 2.37 (0.20) & 51 (96.33) & 7.90 \\
UPK~533 & 267.52 & -9.78 & -7.03 (0.17) & 2.57 (0.16) & 348 (24) & 76.95 & 18.48 (0.34) & 23 (23.52) & 7.90 \\
UPK~535 & 267.66 & -7.22 & -13.08 (0.36) & 3.25 (0.22) & 319 (12) & 81.14 & 63.64 (0.68) & 72 (72.00) & 7.65 \\
UPK~537 & 268.29 & -7.86 & -6.69 (0.17) & 7.58 (0.18) & 603 (37) & 60.17 & 29.99 (0.28) & 42 (68.66) & 7.60 \\
UPK~540 & 270.63 & -17.08 & -4.82 (0.22) & 7.66 (0.19) & 372 (23) & 74.66 & 33.24 (0.04) & 44 (44.99) & 7.50 \\
UPK~542 & 272.86 & -5.81 & -4.87 (0.07) & 2.56 (0.12) & 835 (30) & 50.70 & 21.67 (0.05) & 19 (35.93) & 8.80 \\
UPK~545 & 275.19 & -9.96 & -8.74 (0.33) & 2.77 (0.32) & 331 (28) & 99.33 & 23.37 (0.16) & 142 (142.00) & 7.95 \\
UPK~549 & 276.79 & 0.70 & -1.69 (0.13) & -0.43 (0.11) & 1010 (25) & 24.95 & 6.96 (0.04) & 104 (222.72) & 9.05 \\
UPK~552 & 280.22 & 9.47 & -22.45 (0.38) & 6.31 (0.32) & 345 (61) & 85.98 & 17.85 (0.05) & 43 (43.00) & 7.95 \\
UPK~562 & 288.58 & 2.88 & -11.67 (0.15) & 4.85 (0.16) & 816 (29) & 44.23 & 12.06 (0.09) & 45 (73.57) & 7.90 \\
UPK~567 & 291.21 & -7.60 & -6.37 (0.11) & 0.11 (0.14) & 638 (24) & 63.52 & 12.18 (0.08) & 54 (74.96) & 8.70 \\
UPK~569 & 294.78 & -10.32 & -25.75 (0.63) & -1.05 (0.81) & 252 (26) & 88.53 & 42.96 (0.32) & 63 (63.00) & 7.55 \\
UPK~578 & 302.97 & 5.49 & -12.93 (0.08) & -4.67 (0.08) & 942 (24) & 16.03 & 3.99 (0.02) & 23 (92.38) & 8.85 \\
UPK~579 & 303.01 & 12.41 & -10.73 (0.21) & -3.20 (0.18) & 717 (39) & 101.05 & 36.49 (0.17) & 76 (105.54) & 7.90 \\
UPK~585 & 307.83 & -13.76 & -6.80 (0.15) & -5.55 (0.16) & 710 (44) & 38.48 & 3.84 (0.26) & 27 (44.36) & 7.90 \\
UPK~587 & 308.37 & 12.70 & -8.82 (0.21) & -2.60 (0.21) & 891 (41) & 73.13 & 21.89 (0.17) & 78 (147.19) & 7.95 \\
UPK~594 & 312.49 & -7.05 & -1.98 (0.20) & -3.16 (0.20) & 939 (43) & 69.82 & 22.84 (0.08) & 78 (147.33) & 7.90 \\
UPK~595 & 312.66 & -6.12 & -5.93 (0.14) & -3.92 (0.18) & 875 (74) & 67.38 & 12.59 (0.20) & 56 (105.89) & 8.80 \\
UPK~596 & 313.12 & -11.39 & -4.40 (0.15) & -3.70 (0.19) & 711 (49) & 76.17 & 19.08 (0.12) & 36 (58.90) & 8.00 \\
UPK~599 & 314.03 & 23.16 & -9.95 (0.27) & 0.30 (0.22) & 670 (38) & 74.25 & 18.30 (0.10) & 56 (91.44) & 7.30 \\
UPK~602 & 316.71 & -6.50 & -1.32 (0.21) & -3.94 (0.26) & 886 (54) & 93.77 & 65.51 (0.80) & 131 (245.44) & 7.90 \\
UPK~604 & 318.25 & -0.64 & -4.52 (0.11) & -3.62 (0.11) & 761 (31) & 32.32 & 7.14 (0.10) & 31 (75.21) & 7.15 \\
UPK~605 & 318.84 & -8.37 & -3.63 (0.08) & -4.28 (0.08) & 736 (17) & 22.23 & 3.73 (0.09) & 34 (55.70) & 7.85 \\
UPK~606 & 319.09 & 13.62 & -20.26 (0.40) & -16.80 (0.41) & 170 (15) & 81.53 & 17.63 (0.11) & 40 (40.00) & 7.10 \\
UPK~607 & 319.74 & 1.46 & -4.26 (0.23) & -3.24 (0.23) & 900 (65) & 20.58 & 4.30 (0.06) & 47 (114.21) & 7.10 \\
UPK~612 & 322.59 & -15.03 & -13.27 (0.92) & -12.06 (0.91) & 241 (32) & 181.72 & 79.39 (0.23) & 181 (181.00) & 8.00 \\
UPK~613 & 324.23 & 8.11 & -4.54 (0.09) & -3.63 (0.10) & 871 (32) & 19.00 & 8.20 (0.07) & 18 (43.94) & 7.85 \\
UPK~614 & 324.35 & 2.36 & -1.84 (0.10) & -4.48 (0.07) & 889 (29) & 24.72 & 4.01 (0.14) & 27 (51.15) & 7.95 \\
UPK~617 & 325.25 & 0.04 & -2.71 (0.13) & -5.30 (0.10) & 723 (17) & 21.41 & 5.47 (0.02) & 44 (72.08) & 7.90 \\
UPK~621 & 326.93 & -0.03 & -2.57 (0.29) & -3.08 (0.24) & 866 (44) & 37.68 & 6.88 (0.28) & 87 (164.05) & 8.80 \\
UPK~624 & 329.08 & -10.65 & -1.44 (0.25) & -14.33 (0.28) & 325 (38) & 52.10 & 17.34 (0.13) & 32 (32.00) & 7.65 \\
UPK~626 & 330.69 & 2.09 & -7.30 (0.22) & -5.67 (0.19) & 510 (34) & 40.76 & 9.51 (0.04) & 38 (52.79) & 8.20 \\
UPK~627 & 331.53 & -2.29 & -2.44 (0.16) & -10.06 (0.15) & 530 (25) & 44.69 & 13.57 (0.11) & 34 (40.24) & 8.00 \\
UPK~629 & 332.99 & -8.44 & 0.80 (0.10) & -3.80 (0.11) & 922 (34) & 40.86 & 19.18 (0.18) & 33 (70.92) & 7.90 \\
UPK~630 & 334.33 & -2.71 & -1.75 (0.09) & -2.30 (0.08) & 931 (27) & 12.32 & 4.04 (0.01) & 32 (89.29) & 7.85 \\
UPK~639 & 343.41 & -6.29 & 1.15 (0.28) & -9.30 (0.29) & 707 (59) & 118.23 & 62.90 (0.82) & 135 (187.40) & 8.90 \\
UPK~640 & 343.55 & 4.64 & -11.89 (0.49) & -21.15 (0.47) & 177 (7) & 116.15 & 41.19 (0.10) & 340 (340.00) & 7.50 \\
UPK~642 & 345.57 & -5.90 & 2.31 (0.11) & -4.29 (0.12) & 950 (46) & 42.53 & 21.02 (0.69) & 34 (64.11) & 8.00 \\
UPK~644 & 349.15 & 10.62 & -3.94 (0.16) & -3.46 (0.13) & 742 (29) & 37.99 & 6.87 (0.11) & 42 (68.61) & 7.95 \\
UPK~645 & 349.54 & -9.55 & 2.76 (0.13) & -2.85 (0.12) & 683 (27) & 26.87 & 8.13 (0.03) & 39 (63.89) & 7.95 \\
UPK~654 & 359.58 & -10.83 & 1.13 (0.20) & -8.65 (0.17) & 503 (29) & 33.74 & 11.02 (0.05) & 49 (58.08) & 8.00 \\
UPK~655 & 359.89 & -17.71 & 4.48 (0.77) & -27.20 (0.82) & 153 (5) & 26.24 & 9.83 (0.06) & 35 (35.00) & 7.10 \\
\bottomrule
\end{tabular}
\tabnote{
*UPK stands for the names of the authors' institutions: Ulsan Science High School, Pusan National University, Korea Astronomy and Space Science Institute. Numbers in  parenthesis are errors. \\ 
$^{\rm a}$GMM cluster radius. \\ 
$^{\rm b}$King profile core radius. \\ 
$^{\rm c}$Number of stars within the best-fit ellipse. Numbers in parenthesis are numbers of member stars corrected for the incompleteness due to the magnitude cut at $G=18$.
}
\vskip1\baselineskip
\end{table*}

\subsection{Properties of New Clusters}

Figure~\ref{fig:l_v} shows the distribution of newly discovered open clusters together with known clusters in Galactic coordinates.  The new cluster are less concentrated toward the Galactic plane than the open clusters in MWSC. There are only five clusters at high Galactic latitude with $|b| > 20^{\circ}$. The absence of high latitude new clusters is consistent with their young to intermediate ages, which suggests that they have not yet migrated far from their birth places in the Galactic disk. There are slightly more new open clusters in  $0^{\circ}  <  l  < 180^{\circ}$ than in $180^{\circ}  <  l  < 360^{\circ}$.  There seems to be no significant difference between the Galactic northern sky ($b > 0^{\circ}$) and southern sky ($b < 0^{\circ}$).

\begin{figure}[!t]
	\centering
	\includegraphics[width=1\columnwidth]{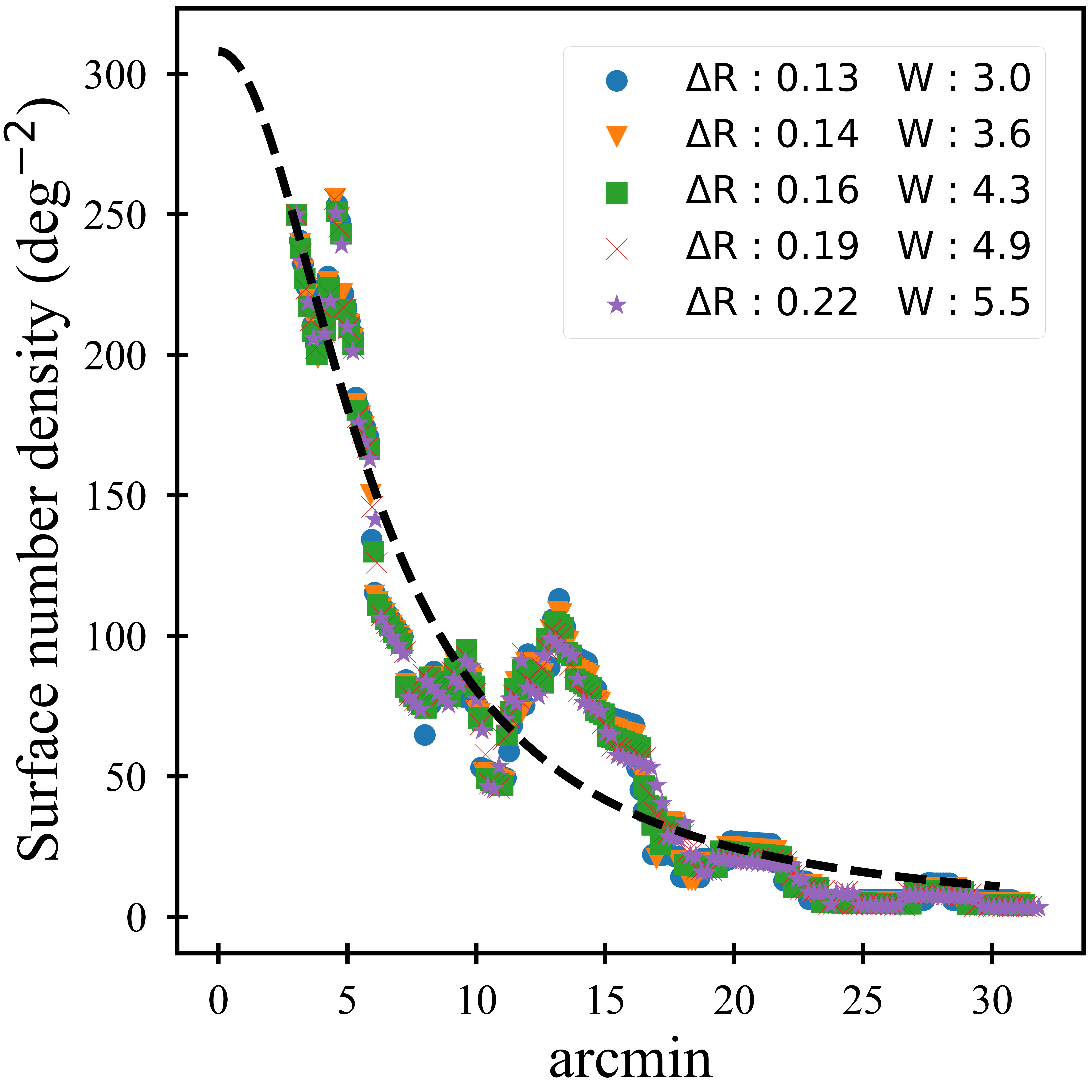}
	\caption{Surface density profiles of UPK~007.  Profiles denoted by different symbols are extracted using different sets of aperture parameters given in the upper right corner; $\Delta R$ is the aperture interval and $W$ is the width of the outermost aperture. The King model  derived from the mean values of $f_{0}$ and $r_{c}$ from the five profiles is indicated by the solid line. }
	\label{fig:kingpf7}
\end{figure}

We plotted the age distribution of the newly found open clusters in Figure~\ref{fig:Hage} together with that of known clusters for comparison. We note that the accuracy of cluster ages is limited because we derived them using a fixed metal abundance of $Z=0.02$. Nevertheless, we can divide the clusters into three age groups: young ($ < 10^{8}$ yr), intermediate ($10^{8} \le$ t (yr) $< 10^{9}$ ), and old ($\ge 10^{9}$ yr).
As shown in Figure~\ref{fig:Hage}, the newly found open clusters are mostly young and intermediate-age clusters that have ages of $7.5\times10^{7}$ yr $\sim$ $10^{9}$ yr. Very young clusters ($<10^{7}$ yr) are slightly more abundant than old clusters. The age distribution of newly found clusters is similiar to that of the known clusters. The age distribution we find is similar to that observed in known clusters by \citet{pis18}. The small number of old open clusters among the newly found open clusters is due to the fact that well defined old open clusters are easy to find because most of them are populous and located at high Galactic
latitudes. Moreover, old open clusters with small numbers of member stars are prone to dissolution by dynamical evolution which drives mass segregation and evaporation of low mass stars \citep{mic64, ter87, fri95}.

\begin{figure}[!t]
	\centering
	\includegraphics[width=1\columnwidth]{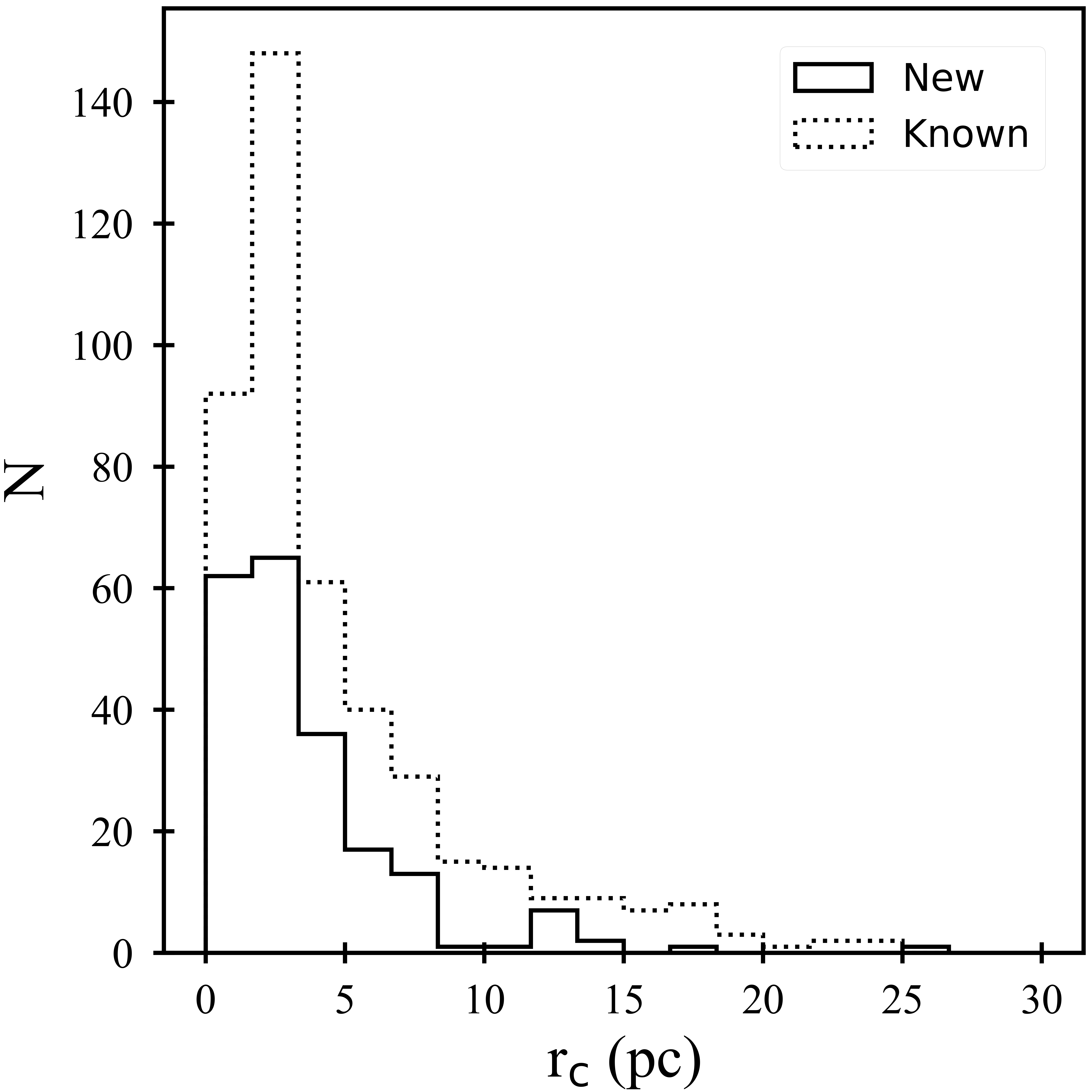}
	\caption{Frequency distribution of core radii ($r_{\rm{c}}$) of our 653
open clusters. Solid lines represent the 207 newly discovered clusters, dotted lines represent the 453 known clusters.}
	\label{fig:sim8}
\end{figure}

Figure~\ref{fig:kingpf7} shows the surface density profile of the newly discovered cluster UPK~007, derived by counting the number of stars in concentric
annuli and division by the annulus area.
We plotted five profiles which were derived using different sets of aperture interval and width. We set the width of the first aperture to half of $r_{\rm{GMM}}$ and decrease the width with increasing distance from the center. The typical width of the outermost aperture is one tenth of $r_{\rm{GMM}}$ which is assumed to be the cluster radius. The averaged surface density profile was fitted  with the King model \citep{king66}
\[
f (r) = \frac{f(0)}{1 + (r/r_{\rm{c}})^{2}} + f_{b}
\]
where $f(0)$ is the central surface density, $r_{\rm{c}}$ is the core radius, and $f_{b}$ is the constant background surface density. We present
the frequency distribution of core radii ($r_{\rm{c}}$) of new open clusters in Figure~\ref{fig:sim8} together with that of the known open clusters.  We estimated the errors in $r_{\rm{c}}$ from the  errors caused by the uncertainty in the derivation of the surface density profile.  We derived five values of $r_{\rm{c}}$ and calculated the mean and standard deviation. We considered the standard deviation of the five measures of  $r_{\rm{c}}$ to be the error in $r_{\rm{c}}$.

\begin{figure}[!t]
	\centering
	\includegraphics[width=1\columnwidth]{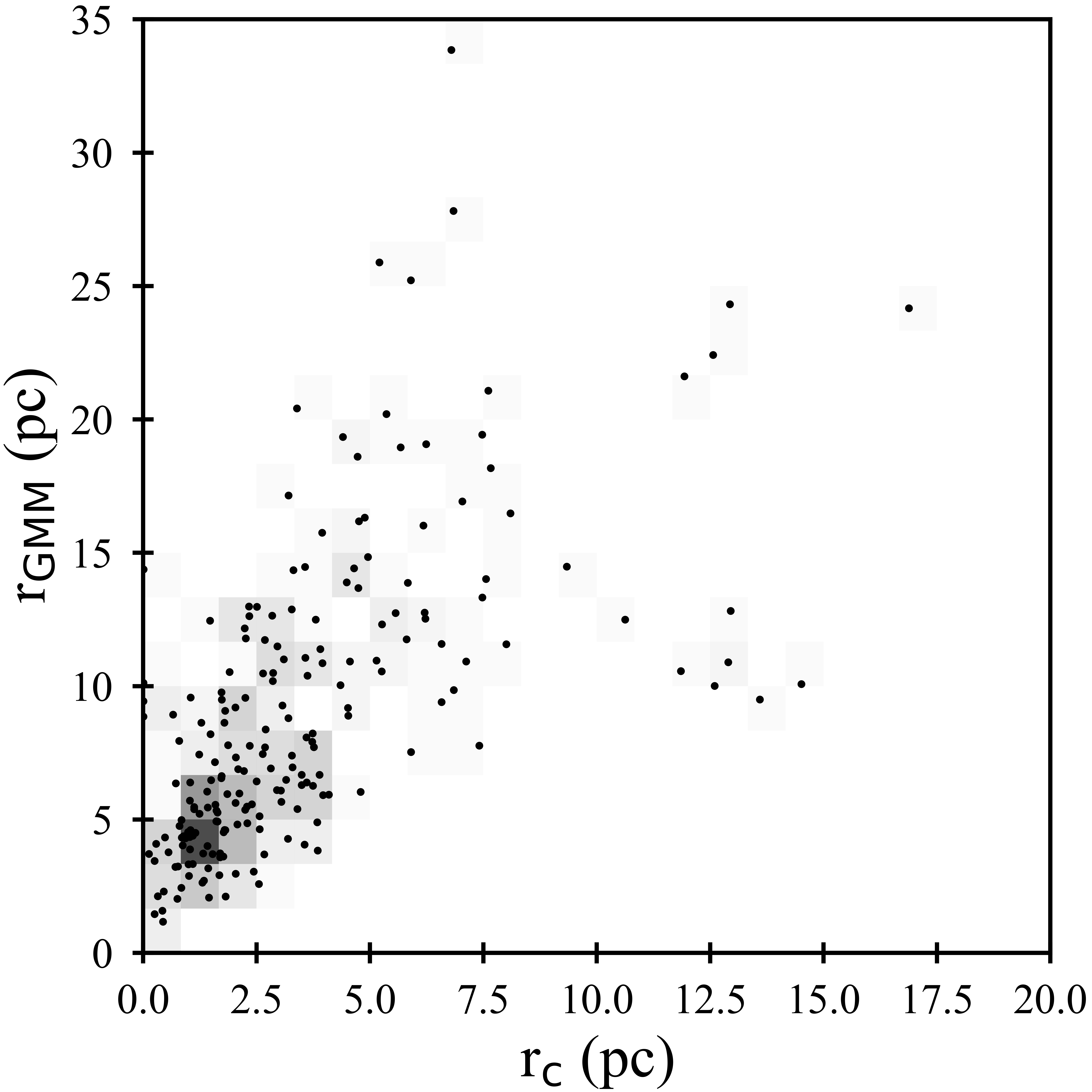}
	\caption{Comparison of $r_{\rm{c}}$ and $r_{\rm{GMM}}$ for our 207 new open clusters.}
	\label{fig:Rgmm2Rc}
\end{figure}

As shown in Figure~\ref{fig:sim8}, most of the new clusters have $r_{\rm{c}}$ smaller than $\sim4$ pc with the average value being 3.7 pc. This is about twice as much as the mean value of 1.8 pc of the clusters in MWSC. The large mean $r_{\rm{c}}$ of the new clusters seems to be due to a more complete census of member stars. The \emph{Gaia} proper motions make it possible to catch escaping cluster stars which are located close to the tidal radius, which is about the same as $r_{\rm{GMM}}$.  As shown in Figure~\ref{fig:Rgmm2Rc}, there is a good correlation between $r_{\rm{GMM}}$ and $r_{\rm{c}}$.
The size of $r_{\rm{c}}$ is about 4 times smaller than $r_{\rm{GMM}}$.  We over plotted the density of clusters in $r_{\rm{GMM}}$- $r_{\rm{c}}$ space to see the number distribution together.

\begin{figure}[!t]
	\centering
	\includegraphics[width=1\columnwidth]{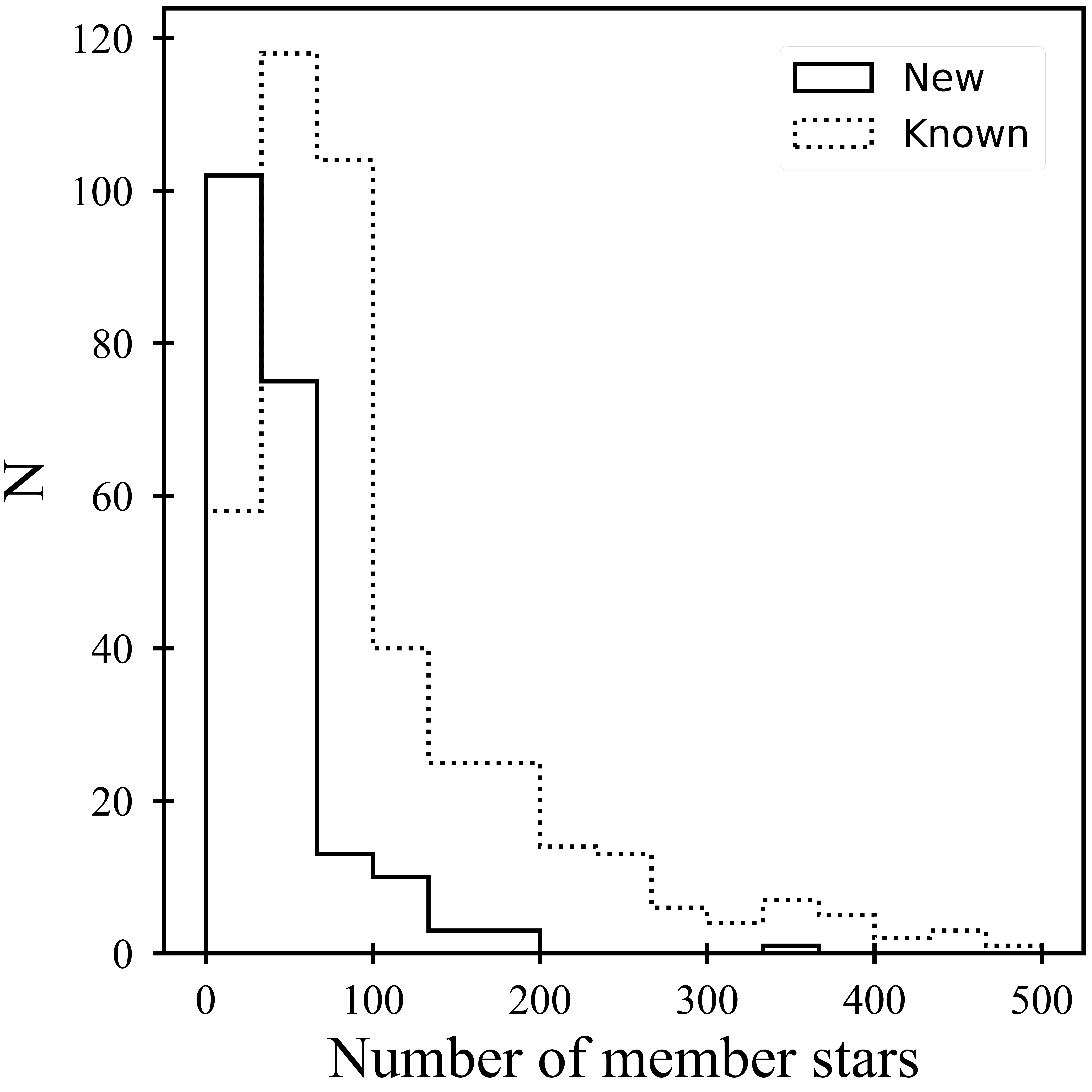}
	\caption{Frequency distribution of numbers of member stars. Solid lines represent newly discovered clusters, dotted lines represent known clusters.}
	\label{fig:sim10}
\end{figure}

\begin{figure}[!t]
	\centering
	\includegraphics[width=1\columnwidth]{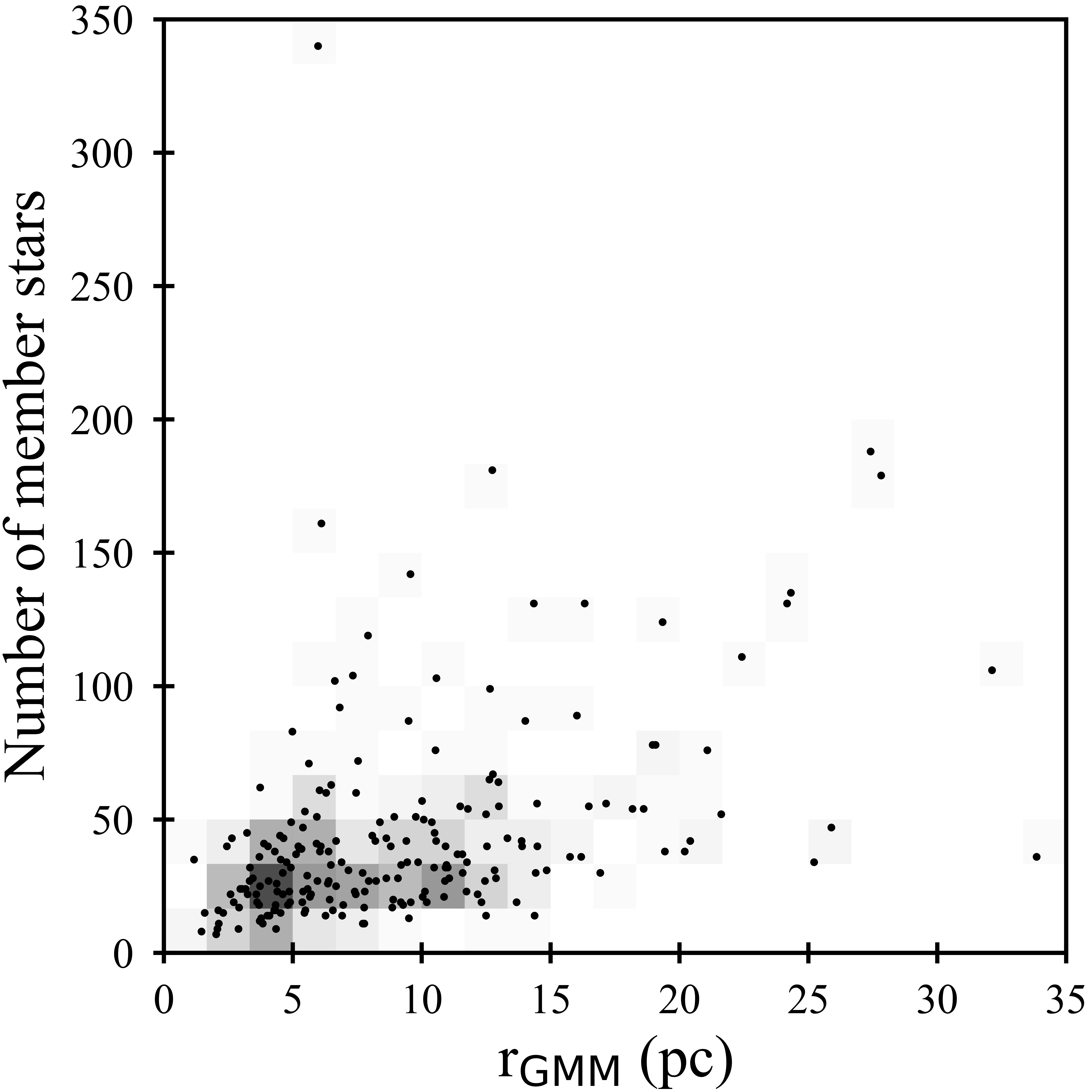}
	\caption{Number of member stars as a function of $r_{\rm{GMM}}$. }
	\label{fig:sim11}
\end{figure}

Figure~\ref{fig:sim10} shows the distribution of the number of members within $r_{\rm{GMM}}$ for our 653 detected open clusters, corrected for the incompleteness caused by the magnitude cut at $G = 18$. We used the luminosity functions derived from $Z =0.02$ PARSEC isochrones \citep{bre12} which assumed the initial mass function of \citet{kro01, kro02}. The correction factor  for the incompleteness $f$ is given by
\[
f={{\int\limits_{M_{l, model}}^{M_{u, obs}} \Psi (M)\,\mathrm{d}M} \over {\int\limits_{M_{l, obs}}^{M_{u, obs}} \Psi (M)\,\mathrm{d}M}}
\]
where $\Psi (M)$ is the luminosity function of the model cluster and $M_{l, model}$ is the  lower limiting magnitude of the model cluster. $M_{l, obs}$ and $M_{u, obs}$ are the lower and upper limitting magnitudes of the observed cluster, respectively.
The majority ($75\%$) of newly discovered clusters have less than 50 member stars while only $27\%$ of the known clusters have that few members. This is the reason why they remained undetected. The most populous cluster (UPK 638) has 340 stars. The second-most populous cluster (UPK 045) has a very elongated shape and is located at $\sim 900$ pc. As shown in Figure~\ref{fig:sim11}, some clusters have unusually small numbers of member stars given their radii.

\section{Summary and Discussion}
\label{sec:sumdis}

We visually found 655 clusters within 1 kpc from the Sun using the astrometric (proper motion, position, parallax) and photometric data of \emph{Gaia} DR2, exploiting the fact that stars bound together in a cluster have similar proper motions and are spatially clustered.  We used only stars brighter than $G$=18 to reduce contamination by field stars caused by the fact that measurement errors increase as the brightness of stars decreases.

We applied mean-shift and GMM analysis to identify clusters in proper motion space and spatial distributions. We derived cluster parameters such as the radii in which cluster stars are located in proper motion space ($r_{\rm{PM}}$) and $l-b$ space ($r_{\rm{GMM}}$) iteratively. We validated cluster candidates by comparing their CMDs with PARSEC $Z=0.02$ isochrones, which resulted in confirmation of 653 open clusters from the 655 visually identified cluster candidates.
We cross-matched all 655 visually identified cluster candidates with previous catalogs. We found 207 new open clusters among the 653 clusters that pass CMD validation. Some of the new open clusters might actually be
associations or small stellar aggregations given the small numbers of member stars.

We present the physical parameters of the 207 new clusters in a catalog that provides Galactic coordinates, mean proper motions, parallaxes, $r_{\rm{GMM}}$, $r_{\rm{c}}$,  numbers of member stars within $r_{\rm{GMM}}$, and ages. We also present properties of our 655 visually selected cluster candidates in an online catalog.  The newly found open clusters have physical properties similar to those of known clusters, with some notable exceptions. Their mean core radius ($r_{\rm{c}}$) is 3.7 pc while that of known clusters is 1.8 pc. Since known clusters are generally more populous than the new open clusters, it is apparent that the known open clusters were the ones that could be found more easily. The majority of newly found clusters have less than $\sim50$ member stars with the largest one having 340. The large core radii and small numbers of member stars of the new clusters are likely due to dynamical evolution that leads to the evaporation of low mass stars from the cluster. The relatively low central concentrations and small numbers of member stars of the new clusters caused their late detection.

The age distribution of the newly found clusters shows that the majority of them are intermediate-age open clusters.  We found that some clusters have a significant number of member stars far from the center of the cluster.  These clusters are thought to have extended halos or stars escaping from their gravitational potentials. There are a number of clusters that have elongated shapes. Some of them show multiple parallax peaks which suggest multiple components. However, the reason for the elongagtion is not clear. Dynamical evolution may play a critical role since most of these clusters are intermediate-age clusters. If the elongated shape  originated from their formation process, we expect to find more elongated structures in younger clusters. Dynamical modeling is required to understand the origin of the elongated morphology.

\acknowledgments

We are grateful to the anonymous referees for valuable comments and suggestions.
This work has made use of data from the European Space Agency (ESA) mission
{\it Gaia} (\url{https://www.cosmos.esa.int/gaia}), processed by the {\it Gaia}
Data Processing and Analysis Consortium (DPAC,
\url{https://www.cosmos.esa.int/web/gaia/dpac/consortium}). Funding for the DPAC
has been provided by national institutions, in particular the institutions
participating in the {\it Gaia} Multilateral Agreement. We also thank the Ulsan Science High School (USHS) for financial support. SGH would like to thank USHS science and mathematics teachers for help with understanding the theoretical background, especially Park Joo-Hyun. This work was partially supported by the 2018 Science and Technology Promotion Fund and Lottery Fund through KOFAC.

\end{document}